\documentclass[12pt]{iopart}
\usepackage{graphicx}
\begin{document}
\title{Circular Geodesics in the Kerr-Newman-Taub-NUT Space-time}
\author{Parthapratim Pradhan}
\address{ Department of Physics, Vivekananda Satabarshiki Mahavidyalaya
(Affiliated to Vidyasagar University), Manikpara, Jhargram, West Midnapur,
West Bengal~721513, India}
\ead{pppradhan77@gmail.com}

\begin{abstract}
In this paper we investigate the equatorial causal (time-like and null)
circular geodesics of the Kerr-Newman-Taub-NUT(Newman-Unti-Tamburino)
black hole in four dimensional Lorentzian geometry. The special characteristics of
this black hole is that it is of \emph{Petrov-Pirani type D} and
the photon trajectories are \emph{doubly degenerate principal null congruence}.
We derive the conditions for existence of innermost stable circular orbit,
marginally bound circular orbit and  circular photon orbit in the background of
Kerr-Newman-Taub-NUT(KNTN) space-time. The effective potential for both
time-like case and null cases have been studied. It is shown that the
\emph{presence of the NUT parameter deforms the shape of the effective potential in contrast with the zero
NUT parameter}.  We further investigate the energy extraction by the Penrose process for
this space-time. It is shown that the efficiency of this black hole depends on both the charge
and NUT parameter. It is observed that the energy gain is maximum when NUT parameter goes to
zero value and for the maximum spin value. When the value of \emph{NUT parameter is increasing
the energy-gain is decreasing}.
\end{abstract}
\maketitle
\section{Introduction}
Although there is as yet no certain observational evidence  for the existence of gravitomagnetic mass or dual
(or magnetic) mass or gravitomagnetic monopole, the investigation of the geodesic properties of KNTN space-time
has great significance from theoretical and conceptual points of view. The  dual mass has also an intrinsic
properties of the space-time in gravitational physics. The possibilities of the observational search for
the gravito-magnetic monopole was first proposed by Lynden-Bell and Nouri-Zonoz\cite{bell}
in 1998. It has been also suggested that signatures of such spacetimes might be found in the spectra of supernovae,
quasars, or active galactic nuclei. In \cite{gonzales} the authors pointed out that zero charge KNTN BH possesses
thin accretion disk which is of astrophysically has an importance because they could be used as models for certain
galaxies, superposition of a black holes and a galaxy or an accretion disk as in the case of quasars.

It is well known that most astro-physical compact objects possess only a small net charge or no charge at
all\cite{ruffini}. Therefore from the astrophysical perspectives, the importance of such studies have very limited
applications. For instance, Kibble in 1980 suggested that gravitomagnetic monopoles would be a natural consequence of
the Big Bang\cite{kibble}. In case of Taub-NUT space-time, the presence of the NUT parameter i.e. gravitomagnetic
monopole produced the lensing pattern and all the geodesics including the null geodesics are lie on a cone\cite{bell97}.
It has been also observed there that there is an extra shear due to the presence of the gravitomagnetic field of
NUT space, which shears the shape of the source. Since the TN space-time possesses gravitomagnetic monopole that affect
the structure of the accretion disk and might offer novel observational prospects\cite{liu,chur}. When we added
the charge parameter with KTN black hole and the study of circular geodesic motion of a neutral test particle in
this gravitational field  has a significance purely from the theoretical point of view.

In Einstein's general relativity, circular geodesics of arbitrary radii are not possible, there exists a minimum
radius below which no circular orbits are possible.  The conditions for the existence of innermost stable circular
orbit (ISCO), marginally bound circular orbit(MBCO)  and circular photon orbits (CPO) have been considered for
the Schwarzschild black hole \cite{sch}, Reissner Nordstr{\o}m(RN) black-hole \cite{sch}, Kerr black hole \cite{sch},
Kerr-Newman(KN) black hole \cite{dk} and  Kerr-Taub-NUT(KTN) black hole more recently \cite{chur}.

But the properties of equatorial circular geodesic of KNTN space-times
have not been considered in the extant literature. Thus in this work, we wish to study
 the complete geodesic structure of a neutral test particle in the KNTN geometry. This is one
way to study the gravitational field around the black hole in the presence of
both charge and NUT parameter.  We also investigate the conditions for the existence of  ISCO, MBCO and CPO
of KNTN black hole. We also calculate the  important astronomical quantities like Kepler
frequency $\Omega_{0}$, angular momentum $L_{0}$, energy $E_{0}$, rotational velocity $v^{\phi}$ which
is an important tool to study the accretion process in the black hole. We further demonstrate that the
Penrose process which had not been investigated previously for the said black hole in the presence of
both charge ($Q$) and NUT parameter ($n$).

Basically the KNTN space-time is an analytic solution of the vacuum
Einstein-Maxwell equations and due to presence  of the NUT charge which
makes the space-time to be asymptotically non-flat, in contrast with KN black hole.
When the metric is expressed in Boyer-Lindquist
type coordinates, two types of coordinate singularity are manifested.  One occurs
at certain values of the radial coordinate where $g_{rr}$ becomes infinite and corresponds to
bifurcate Killing horizons; the other occurs at $\theta=0, \pi$,
where the determinant of the components of the metric vanishes.

The fact that the delineation of the geodesics exhibits the essential features
of the space-time. Therefore a detailed analysis of the circular geodesics
particularly ISCO, MBCO and CPO  are of great significance. Due to the
inclusion of both NUT and charge parameter what effects are manifested
in the Geodetic precession and the dragging of inertial frames
(Lense-Thirring effect) \cite{cp}, this investigation is also crucial
and might have a link to the studies of the equatorial causal geodesics.

Circular geodesic motion in the equatorial plane  ($\theta=\frac{\pi}{2}$) is of
fundamental importance in black hole accretion disk theory also \cite{abram}. Thus ISCO, MBCO
and CPO are more relevant in this regard. Keplerian circular orbits exist in the region
$r>r_{cpo}$, with $r_{cpo}$ being the circular photon orbit. Bound orbit exists in the
region $r>r_{mbco}$, with $r_{mbco}$ being the marginally bound circular orbit, and stable
orbit exist in the region $r>r_{ISCO}$, with $r_{ISCO}$ being the innermost stable
circular orbit ( also called marginally stable circular orbit).
The location of these radii are calculated in the subsequent section.

Now the manuscript is arranged in the following way. In section \ref{kntnut}, we
describe the basic geometry of the KNTN black hole. In section \ref{ecgkn},
we analyze the equatorial circular geodesics of the KNTN black-hole.
Section \ref{cpokntn} devoted to study the geodesics around the photon
orbit. In section \ref{cto}, we derive the ISCO equation for KNTN
black hole.  In section \ref{avtlo}, we derive the Kepler frequency for
time-like geodesics. Section \ref{avtlo} devoted to study the
marginally bound orbit around the KNTN black hole. In section \ref{pp}, we analyze
the Penrose process due to the presence of both charge and NUT parameter.  Finally
we conclude  the discussion in the section \ref{dis}.

\section{\label{kntnut}The KNTN Geometry:}
(a)\emph{ KNTN metric and its property:}
In Boyer-Lindquist (BL) like spherical coordinates $(t, r, \theta, \phi)$ the KNTN black hole is completely
determined by the  four parameters i.e., the mass  $(M)$,  charge ($Q$), angular momentum ($J=aM$) and NUT
parameter ($n$). Thus the corresponding metric(in units where $G=c=1$) is described by \cite{miller,dad,bini}
\begin{eqnarray}
ds^2 &=& -\frac{\Delta}{\rho^2} \, \left[dt-P d\phi \right]^2+\frac{\sin^2\theta}{\rho^2} \,
\left[(r^2+a^2+n^2) \,d\phi-a dt\right]^2
+\rho^2 \, \left[\frac{dr^2}{\Delta}+d\theta^2\right] ~.\label{nkntn}
\end{eqnarray}
where
\begin{eqnarray}
a &\equiv&\frac{J}{M},\, \rho^2 \equiv r^2+(n+a\cos\theta)^2 \\
\Delta &\equiv& r^2-2Mr+a^2+Q^2-n^2\\
P &\equiv& a\sin^2\theta-2n\cos\theta  ~.\label{adeltap}
\end{eqnarray}
The electromagnetic field 2-form would be given by
$$
F = \frac{Q}{\rho^4}[r^2-(n+a\cos\theta)^2]dr\wedge (dt-P d\phi) +
$$
\begin{eqnarray}
\frac{2aQr\sin\theta\cos\theta}{\rho^4} d\theta \wedge[(r^2+a^2+n^2) \,d\phi-a dt] ~.\label{fieldt}
\end{eqnarray}

Note that when $Q=0$, the electromagnetic field tensor vanishes and the metric satisfies the vacuum
Einstein equations. When $n=0$, the specific geometry reduces to Kerr-Newman geometry \cite{sch} and
when $Q=n=0$, the geometry reduces to Kerr geometry \cite{sch}.

\begin{figure}
\begin{center}
{\includegraphics[width=0.45\textwidth]{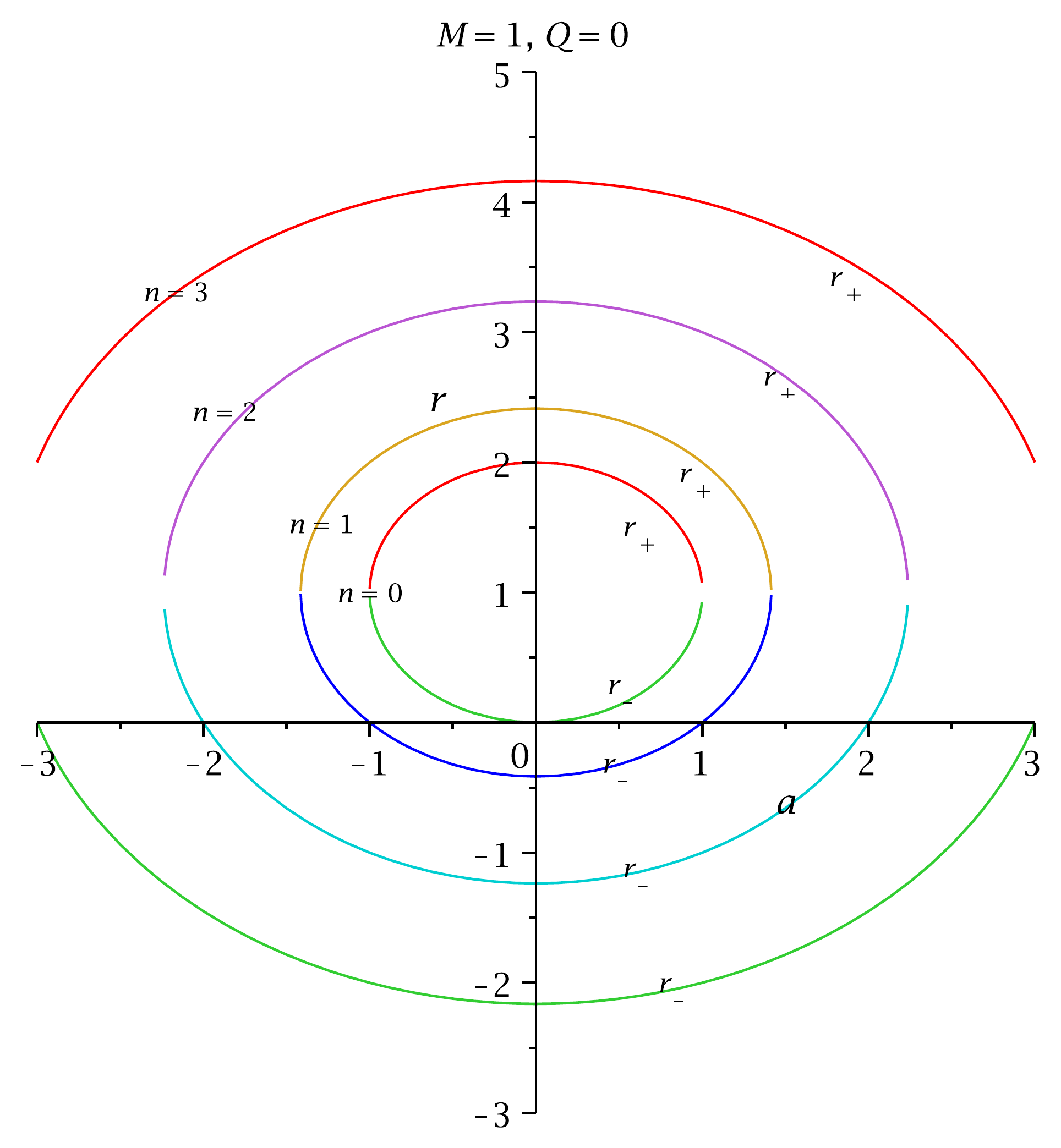}}
{\includegraphics[width=0.45\textwidth]{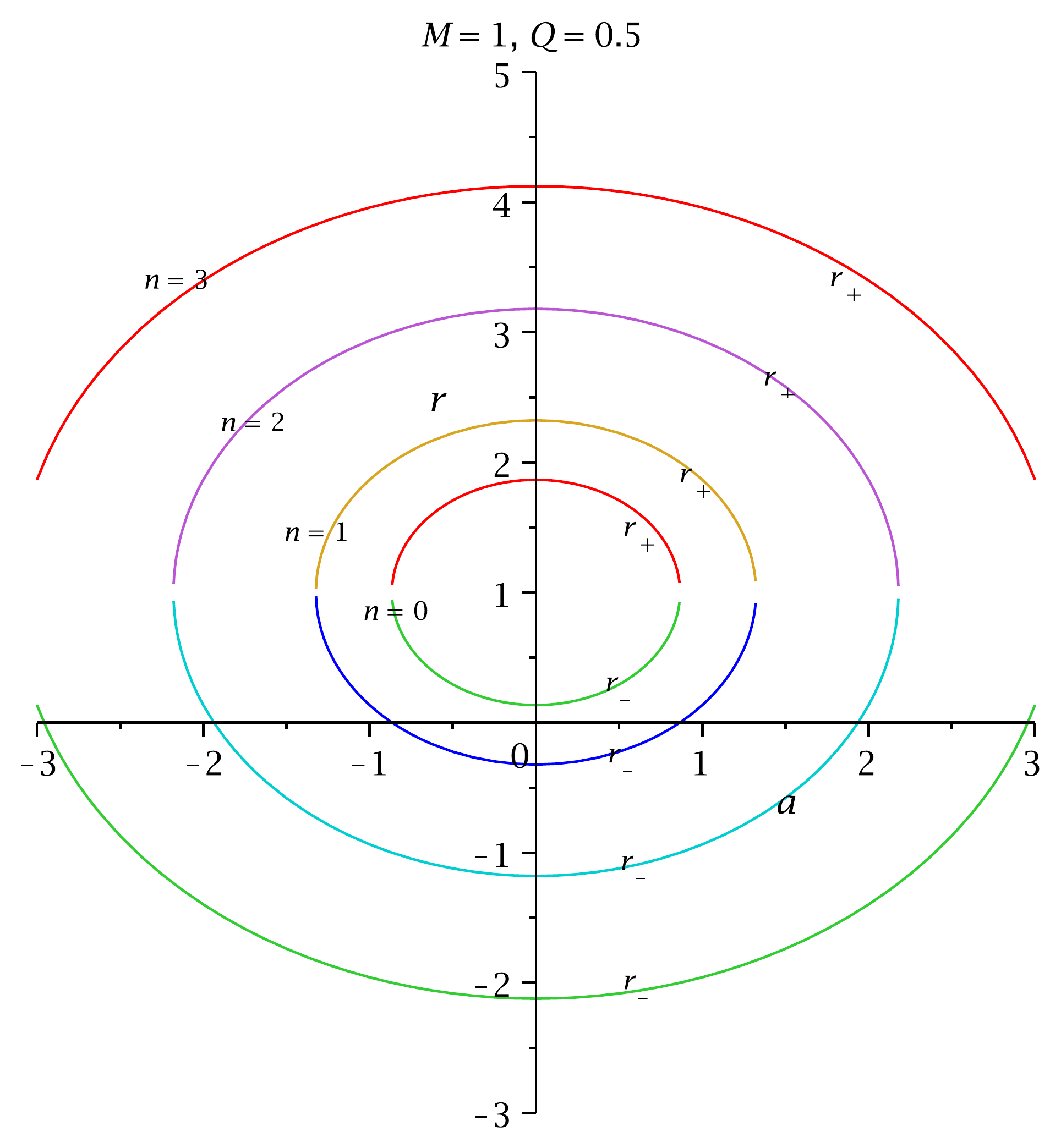}}
\end{center}
\caption{The figure depicts the horizon structure of KTN and KNTN black hole.
\label{hr}}
\end{figure}

The radius of the horizon is determined by the solution of the function $\Delta=0$.
i.e.,
\begin{eqnarray}
r=r_{\pm}\equiv M\pm\sqrt{M^2-a^2-Q^2+n^2}\,\, \mbox{and}\,\,  r_{+}> r_{-}
\end{eqnarray}
where $r_{+}$ is called event horizon (${\cal H}^+$) or outer horizon and $r_{-}$  is called Cauchy
horizon (${\cal H}^{-}$) or inner horizon. From Fig.\ref{hr}, we can see the horizon
structure of Kerr, Kerr-Newman(KN), Kerr-Taub-NUT(KTN) and KNTN black hole . Due to the presence of
NUT parameter the horizon structure of KTN and KNTN black hole are increases in size in
comparison with the Kerr black hole and KN black hole.
%


%
%

%
%

%
%
The static limit surface (outer region of the ergo-sphere)  is  at $g_{tt}=0$ i.e.
\begin{eqnarray}
r=r_{ergo}\equiv M+\sqrt{M^2-a^2\cos^2\theta-Q^2+n^2}\,\, ~.\label{ergokntn}
\end{eqnarray}

\begin{figure}
\begin{center}
{\includegraphics[width=0.45\textwidth]{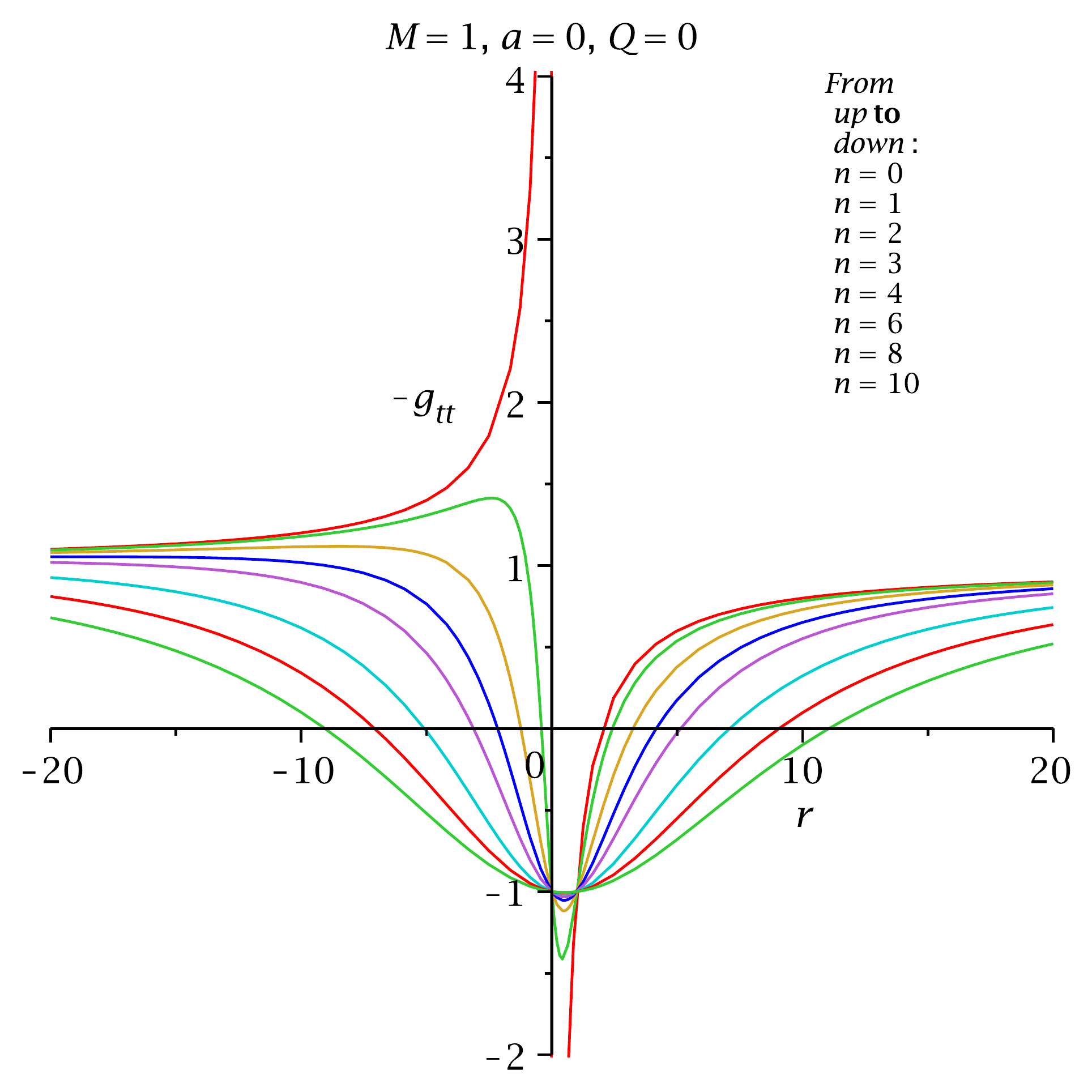}}
{\includegraphics[width=0.45\textwidth]{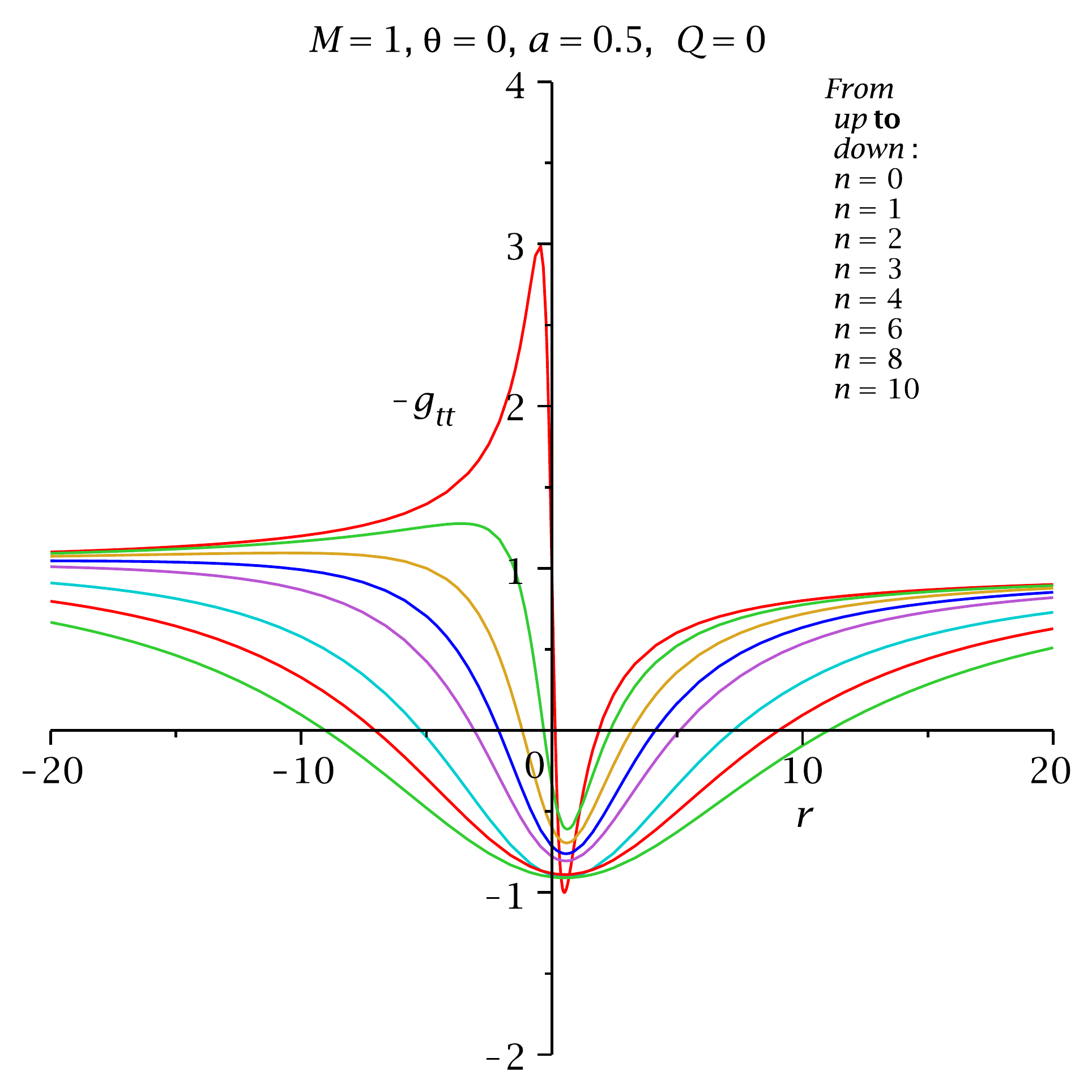}}
\end{center}
\caption{The figure shows the variation  of $-g_{tt}$  with $r$ for TN and KTN black hole.
\label{mf}}
\end{figure}

\begin{figure}
\begin{center}
{\includegraphics[width=0.45\textwidth]{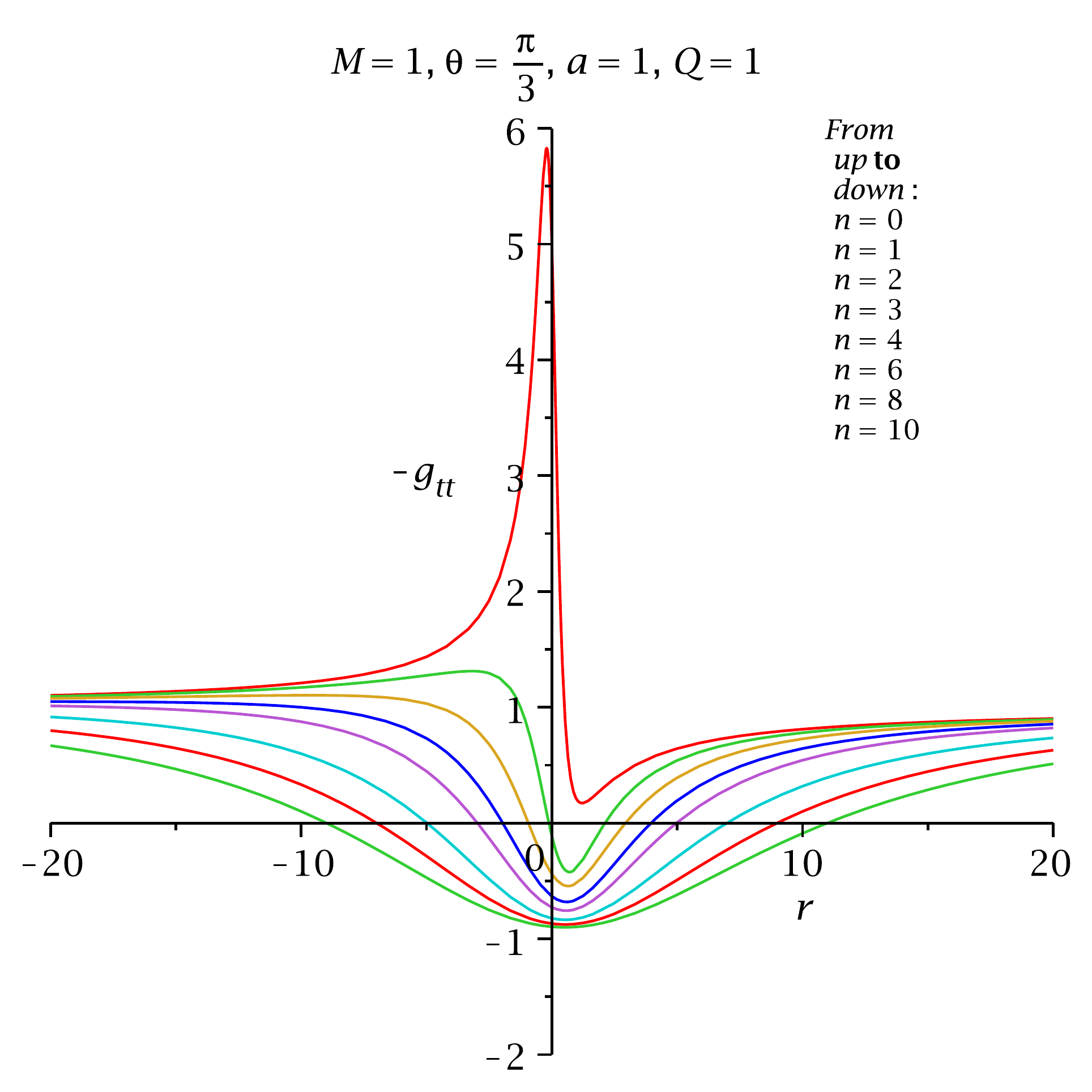}}
{\includegraphics[width=0.45\textwidth]{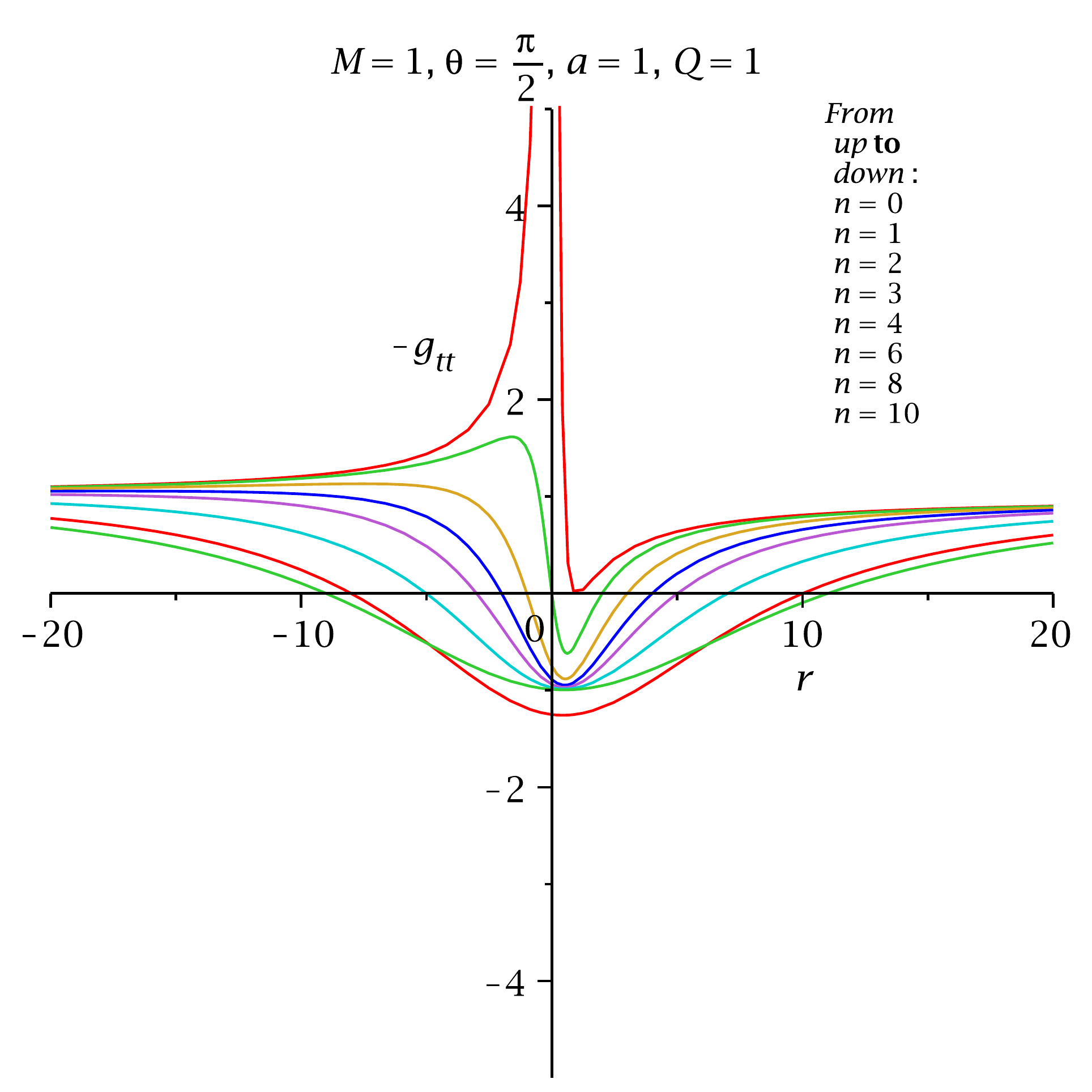}}
\end{center}
\caption{The figure shows the variation  of $-g_{tt}$  with $r$ for KNTN black hole.
\label{mf1}}
\end{figure}

\begin{figure}
\begin{center}
{\includegraphics[width=0.45\textwidth]{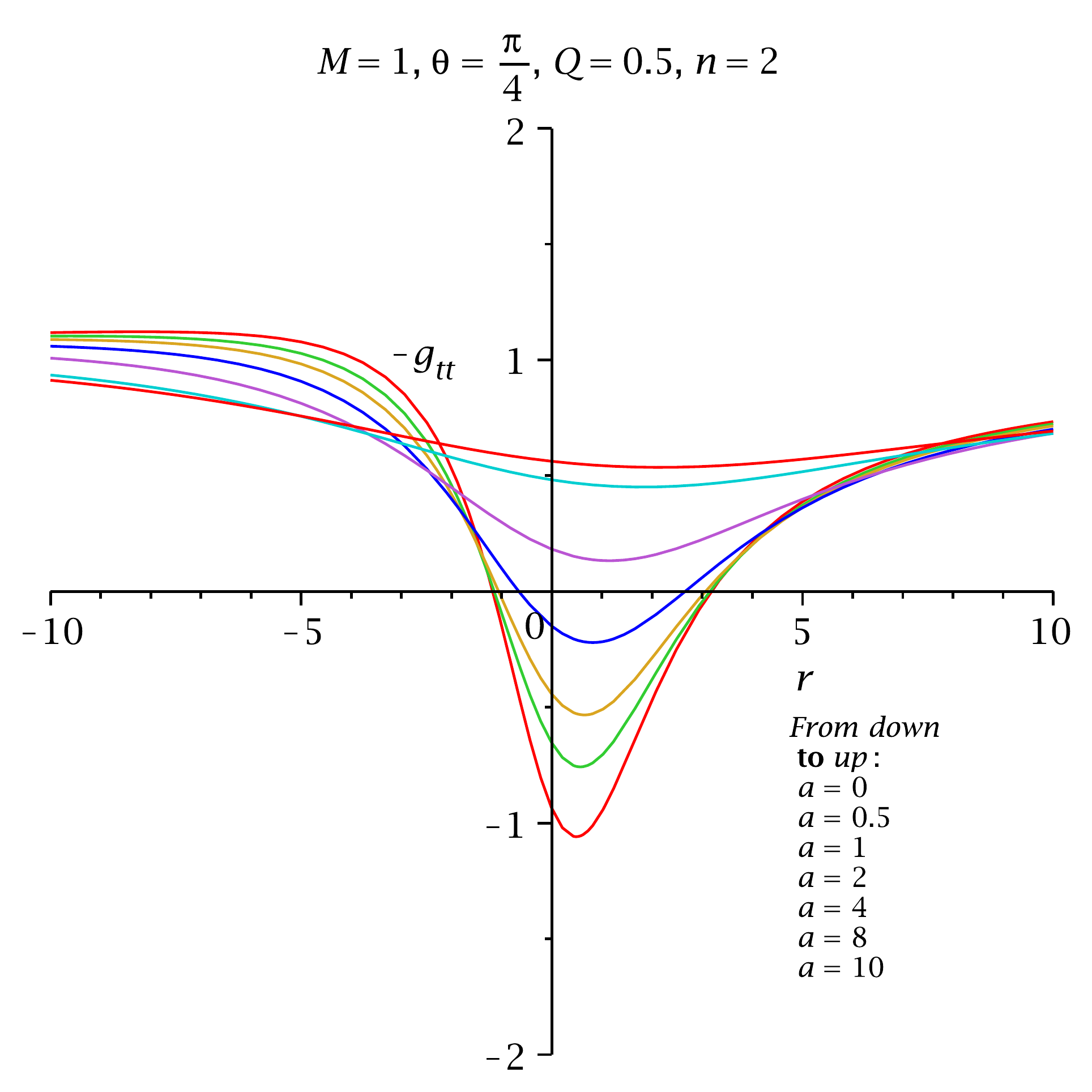}}
{\includegraphics[width=0.45\textwidth]{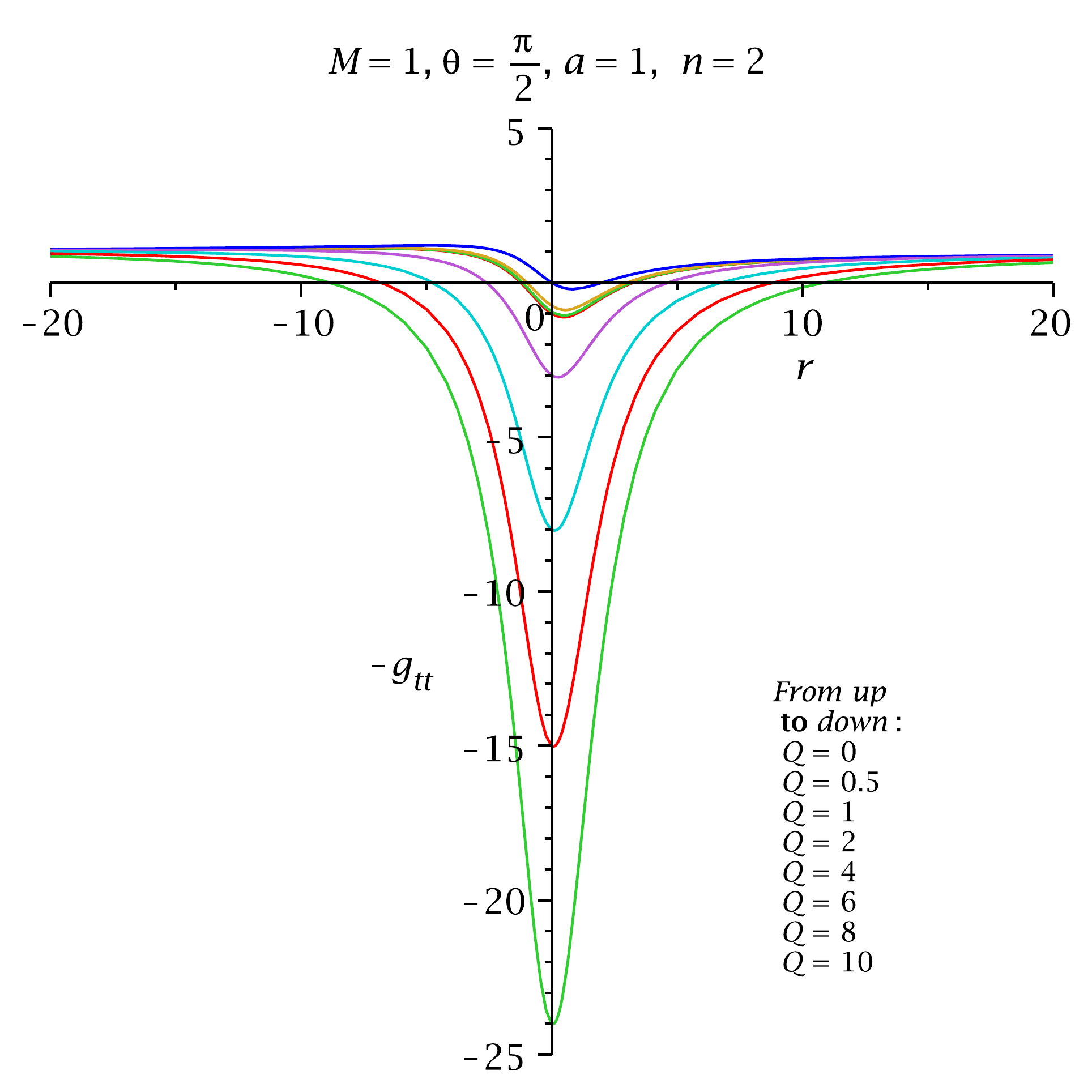}}
\end{center}
\caption{The figure shows the variation  of $-g_{tt}$  with $r$ for KNTN black hole.
\label{mf2}}
\end{figure}

\begin{figure}
\begin{center}
{\includegraphics[width=0.45\textwidth]{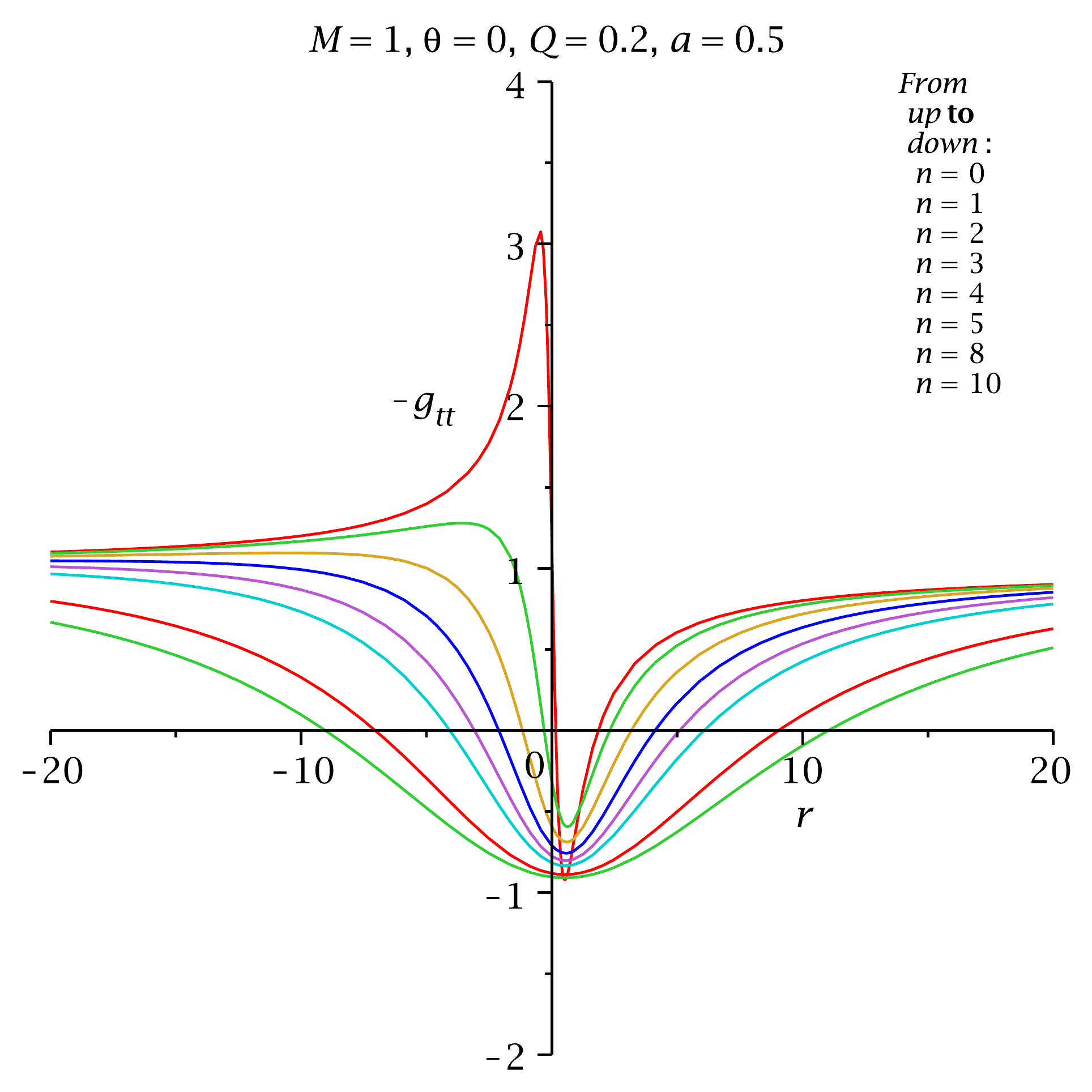}}
\end{center}
\caption{The figure shows the variation  of $-g_{tt}$  with $r$ for KNTN black hole.
\label{mf3}}
\end{figure}

The characteristics of the variation of $-g_{tt}$  with radial coordinate
is shown in Fig.\ref {mf}, Fig. \ref{mf1}, Fig.\ref{mf2} and Fig.\ref{mf3}.
It is observed that the shape of the ergo-region get modified when we incorporated
both charge and NUT parameter in comparison with zero charge and zero NUT parameter.

As long as $M^2+n^2 \geq Q^2 + a^2$ the KNTN metric describes a black hole,
otherwise it has a naked ring-like singularity. When $M^2+n^2 = Q^2 + a^2$, the
situation is called extremal situation in gravitational physics.

The metric components (\ref{nkntn}) of a KNTN space-time are independent of the
BL time coordinate $t$ and angular coordinate $\phi$. This implies that
$\xi \equiv \partial_t$ and $\zeta \equiv \partial_{\phi}$ are Killing
vectors associated  with stationarity  and axial symmetry of the black hole.
Hence, there are two conserved quantities; energy and angular momentum
along the motion of the particle which can be labeled as $E$ and $L$.
The scalar products of these Killing vectors with themselves and each other
are
\begin{eqnarray}
\xi.\xi &\equiv& g_{tt}=-\frac{(\Delta-a^2\sin^2\theta)}{\rho^2} \nonumber\\
\xi.\zeta &\equiv & g_{t\phi}=\frac{2}{\rho^2}[P \Delta-a(\rho^{2}+aP)\sin^2\theta] \nonumber\\
\zeta.\zeta &\equiv& g_{\phi\phi}=\frac{[(\rho^{2}+aP)^{2}\sin^2\theta-\Delta P^{2}]}{\rho^{2}}  ~.\label{gtt}
\end{eqnarray}
These components of the metric gives us physical information about the symmetry
of the space-time. An observer who moves along a geodesics of constant $(r, \theta)$
is called stationary observers whose angular velocity is given by
\begin{eqnarray}
\Omega=\frac{d\phi}{dt}=\frac{u^{\phi}}{u^{t}}=\omega &\equiv& -\frac{g_{t\phi}}{g_{\phi\phi}}
=\frac{2[P \Delta-a(\rho^{2}+aP)\sin^2\theta]}{[(\rho^{2}+aP)^{2}\sin^2\theta-\Delta P^{2}]} ~.\label{omega}
\end{eqnarray}
This is called frame dragging effect which manifested due to the presence of the
off diagonal components of the metric i.e. $g_{t\phi}\neq 0$.


%

%

\emph{(b) Red-Shift factor and Red-Shift :}
The ``red-shift factor'' ($\mathcal{R}$) \cite{mtw} in terms of angular velocity
$\Omega$ for KNTN black hole is given by
\begin{eqnarray}
\mathcal{R} &=& \frac{d\tau}{dt}= \frac{1}{u^{t}}=\sqrt{-[g_{tt}+2\Omega g_{t\phi}
 +\Omega^{2}g_{\phi\phi}]}  ~. \label{redfac}
\end{eqnarray}
The ``red-shift''($\textbf{z}$) \cite{mtw} for KNTN black hole is
\begin{eqnarray}
\textbf{z} &\equiv& \frac{\Delta \lambda}{\lambda}=\frac{\lambda_{received}-\lambda_{emitted}}{\lambda_{emitted}}
=u^{t}-1=\frac{1}{\sqrt{-[g_{tt}+2\Omega g_{t\phi}
 +\Omega^{2}g_{\phi\phi}]}}-1  ~. \label{red}
\end{eqnarray}
where $\Omega =-\frac{g_{t\phi}}{g_{\phi\phi}}$ and the
four velocity components are $u^{\mu}=(u^{t},0,0,\Omega u^{t})$.

The normalization condition of the four velocity $u.u=-1$ shows that
\begin{eqnarray}
u^{t} &=& \frac{1}{\sqrt{-[g_{tt}+2\Omega g_{t\phi}
 +\Omega^{2}g_{\phi\phi}]}}  ~. \label{red1}
\end{eqnarray}
A straightforward calculation shows that
\begin{eqnarray}
\mathcal{R}^{2} &=& \frac{1}{(u^{t})^{2}}= \frac{\Delta \rho^{2}\sin^{2}\theta}
{[(\rho^{2}+aP)^{2}\sin^2\theta-\Delta P^{2}]} ~. \label{redkn}
\end{eqnarray}
As the horizons are approached i.e. $\Delta\rightarrow 0$ the red-shift factor
of both the horizons (${\cal H}^\pm$) are given by
\begin{eqnarray}
\mathcal{R}_{\pm} &=& \frac{\rho_{\pm}\sqrt{\Delta}}{r_{\pm}^{2}+a^{2}+n^{2}}  ~. \label{rsh}
\end{eqnarray}
\emph{(b) Proper Acceleration:}
By computing the magnitude $a=\sqrt{a^{\mu}a_{\mu}}$ of the proper acceleration
of a stationary observer in an orbit of constant angular speed  in the KNTN space-time
we shall find an important properties of both the horizons (${\cal H}^\pm$).

As the horizons (${\cal H}^\pm$) are approached i.e. $(r\rightarrow r_{\pm})$,
the proper acceleration
\footnote{The proper four acceleration can be determine
by using this formula: $a^{\mu}\equiv u^{\nu}\nabla_{\nu}u^{\mu} =[{\Gamma^{\mu}}_{tt}
+2\Omega{\Gamma^{\mu}}_{t\phi}+\Omega^{2}{\Gamma^{\mu}}_{\phi\phi}](u^{t})^{2}$, where we have
used $u^{r}=u^{\theta}=0$ and $u^{t}_{,t}=0$ from the stationarity $(g_{\mu\nu, t}=0)$ of the
geometry of the given metric \cite{bcw,dog}.} of a
stationary observer in the KNTN geometry is found to be
\begin{eqnarray}
a_{\pm} &=& \frac{r_{\pm}-M}{\rho_{\pm}\sqrt{\Delta}}  ~. \label{prac}
\end{eqnarray}
Thus the product of the proper acceleration and the red-shift factor of both
the horizons are :
\begin{eqnarray}
a_{\pm}\mathcal{R}_{\pm} &=& \frac{r_{\pm}-M}{r_{\pm}^{2}+a^{2}+n^{2}}  ~. \label{protar}
\end{eqnarray}
Interestingly this is equal to the surface gravity ${\kappa}_{\pm}$  of both the
horizons ${\cal H}^\pm$ in the KNTN space-time.

Therefore one could find simply
\begin{eqnarray}
a_{\pm}\mathcal{R}_{\pm} &=& \frac{r_{\pm}-M}{r_{\pm}^{2}+a^{2}+n^{2}}= {\kappa}_{\pm} ~. \label{prtar}
\end{eqnarray}
Alternatively one could suggest that the surface gravity of both the horizons
can be considered as the limit of the product of the proper acceleration and the
red-shift factor for a stationary observer .

(d)\emph{ Area, Entropy and Black Hole Temperature:}
Now the area of both the horizons (${\cal H}^\pm$) for KNTN space-times are
\begin{eqnarray}
{\cal A}_{\pm} &=& \int^{2\pi}_0\int^\pi_0 \sqrt{g_{\theta\theta} g_{\phi \phi}}d\theta d\phi
=4\pi(r_{\pm}^2+a^2+n^2) ~.\label{arKNTN}
\end{eqnarray}

The angular velocity of ${\cal H}^\pm$ are
\begin{eqnarray}
 {\Omega}_{\pm} &=& \frac{a}{r_{\pm}^2+a^2+n^2}  ~. \label{omkntn}
\end{eqnarray}
The semiclassical Bekenstein-Hawking entropy of ${\cal H}^\pm$ reads
(in units in which $G=\hbar=c=1$)
\begin{eqnarray}
{\cal S}_{\pm} &=& \frac{{\cal A}_{\pm}}{4} =\pi(r_{\pm}^2+a^2+n^2) ~.\label{etpKntn}
\end{eqnarray}
The surface gravity of ${\cal H}^\pm$ is
\begin{eqnarray}
{\kappa}_{\pm} &=& \frac{r_{\pm}-r_{\mp}}{2(r_{\pm}^2+a^2+n^2)} \,\, \mbox{and}\,\,  \kappa_{+}> \kappa_{-} ~.\label{sgKNtn}
\end{eqnarray}
and the black hole temperature or Hawking temperature of ${\cal H}^\pm$ reads
\begin{eqnarray}
T_{\pm}&=&\frac{{\kappa}_{\pm}}{2\pi} =\frac{r_{\pm}-r_{\mp}}{4\pi (r_{\pm}^2+a^2+n^2)}  ~.\label{tmKNtn}
\end{eqnarray}
It should be noted that event horizon is hotter than the Cauchy horizon i.e. $T_{+} > T_{-} $.

The Komar energy for ${\cal H}^\pm$ is \cite{ppkntn} given by
\begin{eqnarray}
E_{\pm} &=& \pm\sqrt{M^2-a^2-Q^2+n^2} ~. \label{kekntn}
\end{eqnarray}
\section{\label{ecgkn} Equatorial circular geodesics of the KNTN Black-Hole:}
To determine the geodesics in the equatorial plane for the KNTN space-time we shall follow the book
of S. Chandrashekar \cite{sch}. To compute the geodesic motion of a neutral test particle in this plane we
set $\dot{\theta}=0$ and $\theta=constant=\frac{\pi}{2}$.

Thus  the necessary Lagrangian for this motion reads
$$
2{\cal L} = -\left(\frac{r^2-2Mr+Q^2-n^2}{r^2+n^2}\right)\,{\dot{t}}^2
-2a\left(\frac{2Mr-Q^2+n^2}{r^2+n^2}\right)\,\dot{t}\,\dot{\phi}+
$$
\begin{eqnarray}
\frac{r^2+n^2}{\Delta}\,{\dot{r}}^2+\frac{\left[(r^2+n^2)^2+a^2(r^2+n^2)+a^2(2Mr-Q^2+2n^2)\right]}{r^2+n^2}
\,{\dot{\phi}}^2 ~.\label{lag}
\end{eqnarray}
The generalized momenta can be written as
\begin{eqnarray}
p_{t} =-\left(\frac{r^2-2Mr+Q^2-n^2}{r^2+n^2}\right) \,\,\dot{t}
-a\left(\frac{2Mr-Q^2+n^2}{r^2+n^2}\right) \,\dot{\phi}=-E =Const ~.\label{pt}
\end{eqnarray}
$$
p_{\phi} =-a\left(\frac{2Mr-Q^2+n^2}{r^2+n^2}\right)\,\dot{t} +
$$
\begin{eqnarray}
\frac{\left[(r^2+n^2)^2+a^2(r^2+n^2)+a^2(2Mr-Q^2+2n^2)\right]}{r^2+n^2}\,\dot{\phi}=L=Const ~.\label{pphi}
\end{eqnarray}
\begin{eqnarray}
p_{r} = \frac{r^2+n^2}{\Delta}\, \dot{r}  ~.\label{pr}
\end{eqnarray}
Here $(\dot{t},~\dot{r},~\dot{\phi})$ denotes differentiation with respect to affine parameter ($\tau$). Since the Lagrangian
does not depend on `t' and `$\phi$', so $p_{t}$ and $p_{\phi}$ are conserved quantities. The independence of the Lagrangian
on `t' and `$\phi$' manifested, the stationarity and the axi-symmetric character of the KNTN space-time.

Thus the Hamiltonian is given by
\begin{eqnarray}
\cal H &=& p_{t}\,\dot{t}+p_{\phi}\,\dot{\phi}+p_{r}\,\dot{r}-\cal L ~.\label{hamil}
\end{eqnarray}

In terms of the metric the Hamiltonian is given by
$$
2{\cal H} = -\left(\frac{r^2-2Mr+Q^2-n^2}{r^2+n^2}\right) \,\dot{t}^{2}
   -2a \left(\frac{2Mr-Q^2+n^2}{r^2+n^2}\right)\dot{t}\,\dot{\phi}
$$
\begin{eqnarray}
+\frac{r^2+n^2}{\Delta}\dot{r}^2+
  \frac{\left[(r^2+n^2)^2+a^2(r^2+n^2)+a^2(2Mr-Q^2+2n^2)\right]}{r^2+n^2} ~.\label{hamh}
\end{eqnarray}
Since the Hamiltonian is independent of `t', thus we can write,
\begin{eqnarray}
2\cal H &=& -\left[\left(\frac{r^2-2Mr+Q^2-n^2}{r^2+n^2}\right) \,\dot{t}+
 a \left(\frac{2Mr-Q^2+n^2}{r^2+n^2}\right)\right]\,\dot{t}+\frac{r^2+n^2}{\Delta}\,\dot{r}^2
  +\nonumber \\[4mm] &&
  \left[-a\left(\frac{2Mr-Q^2+n^2}{r^2+n^2}\right) \,\dot{t}
+\frac{\left\{(r^2+n^2)^2+a^2(r^2+n^2)+a^2(2Mr-Q^2+2n^2)\right\}}{r^2+n^2} \right]\,\dot{\phi} \nonumber \\
        &=&-E\,\dot{t}+L\,\dot{\phi}+\frac{r^2+n^2}{\Delta}\,\dot{r}^2=\epsilon=const ~.\label{2hd}
\end{eqnarray}
Here $\epsilon=-1$ for time-like geodesics, $\epsilon=0$ for light-like geodesics and $\epsilon=+1$ for
space like geodesics.
Solving equations (\ref{pt}) and (\ref{pphi}) for $\dot{\phi}$ and $\dot{t}$, we get
\begin{eqnarray}
\dot{\phi} &=& \frac{\left[\left(r^2-2Mr+Q^2-n^2\right)L
       +\left(a(2Mr-Q^2+2n^2)\right)E\right]}{(r^2+n^2)\Delta} ~.\label{uphi}\\
\dot{t} &=& \frac{\left[\left((r^2+n^2)(r^2+n^2+a^2)+a^2(2Mr-Q^2+2n^2)\right)E
       -a\left(2Mr-Q^2+2n^2\right)L\right]}{\Delta(r^2+n^2)} ~.\label{ut}
\end{eqnarray}
Putting these values in equation (\ref{2hd}), we get the radial equation
that governing the geodesic motion of the KNTN space-time :
$$
(r^2+n^2)^2\dot{r}^{2} = E^2(r^2+n^2)^2+\left(2Mr-Q^2+2n^2\right)\left(aE-L\right)^2
$$
\begin{eqnarray}
+\left(a^2E^2-L^2\right)(r^2+n^2)+\epsilon \Delta (r^2+n^2) ~.\label{radial}
\end{eqnarray}
Using this radial Eq. (\ref{radial}) with the Eqs. (\ref{uphi}) and (\ref{ut}) for the other components of
the four velocity, we could study many interesting properties
of the orbits of particles and light rays in the equatorial plane. We could also
calculate the radii of circular orbits, the radii of unstable circular photon orbits, etc. These are all different,
depending upon whether the particle or light ray is rotating with the black hole(co-rotating) or in the
opposite direction (counter-rotating).

\section{\label{cpokntn} The Circular Photon Orbit:}
As we have described, $\epsilon=0$ for null geodesics and the radial equation (\ref{radial}) becomes
\begin{eqnarray}
\dot{r}^{2} &=& E^2+\left(\frac{2Mr-Q^2+2n^2}{(r^2+n^2)^2}\right)\left(aE-L\right)^2 +\frac{\left(a^2E^2-L^2\right)}{r^2+n^2}
~.\label{nul}
\end{eqnarray}
To distinguish the geodesics,  it is important to introduce the impact parameter $D=\frac{L}{E}$ rather than by $L$.
(a) \emph{The Special Case $L=aE$:}
When $L=aE$, the impact parameter becomes $D=a$ which play an important role to study the radial geodesics.
Thus the equations (\ref{uphi}, \ref{ut}, \ref{nul}) reduce to
\begin{eqnarray}
\dot{\phi} &=&  \frac{aE}{\Delta} ~.\label{uphir}\\
\dot{t} &=& \frac{(r^2+n^2+a^2)E}{\Delta} ~.\label{utr}\\
\dot{r} &=& \pm E ~.\label{urr}
\end{eqnarray}
Here the dot denote derivative with respect to the affine parameter $\lambda$. Now the
equations governing the $t$ and $\phi$ are
\begin{eqnarray}
\frac{dt}{dr} &=& \pm \frac{(r^2+n^2+a^2)}{\Delta} \,\,\, \mbox{and} \,\,\,
\frac{d\phi}{dr}  = \pm \frac{a}{\Delta} ~.\label{dphidr}
\end{eqnarray}
The solutions of these equations are
\begin{eqnarray}
\pm t &=& r+\frac{r_{+}^2+a^2+n^2}{r_{+}-r_{-}}\ln|\frac{r}{r_{+}}-1|+\frac{r_{-}^2+a^2+n^2}{r_{+}-r_{-}}\ln|\frac{r}{r_{-}}-1|,~
\label{pmt}\\
\pm \phi &=&\frac{a}{r_{+}-r_{-}}\ln|\frac{r}{r_{+}}-1|+\frac{a}{r_{-}-r_{+}}\ln|\frac{r}{r_{-}}-1|,~
\label{pmphi}
\end{eqnarray}
These solutions exhibit the characteristic behaviours of $t$ and $\phi$ blows up  as the extremal limits
are taken. The fact that radial null geodesics  are \emph{independent} of the charge parameter of the
space-time. These radial null geodesics described by the equations (\ref{dphidr}) are members of the
shear-free principal null congruences are confined to the equatorial plane.

Thus the principal null congruences of the KNTN geometry are
\begin{eqnarray}
k^{t} & \equiv & \frac{dt}{d\lambda}=\frac{(r^2+n^2+a^2)E}{\Delta} \\
k^{r} &\equiv &  \frac{dr}{d\lambda}=\pm E ~.\label{ktkr}\\
k^{\theta} & \equiv & \frac{d\theta}{d\lambda}= 0 \\
k^{\phi} & \equiv & \frac{d\phi}{d\lambda}  = \frac{aE}{\Delta} ~.\label{kthkphi}
\end{eqnarray}
Here $+$ for outgoing photon and $-$ for ingoing photon.

The significance of these photon trajectories are that they in fact mold themselves to the space-time
curvature in such a way that, if $C_{abcd}$ is the Weyl conformal tensor and
$\ast  C_{abcd}=\epsilon_{abef} C_{cd}^{ef}$ is its dual then
\begin{eqnarray}
C_{abc[d}k_{\epsilon]}k^{b}k^{c}=0, \,\,
{\ast}C_{abc[d}k_{\epsilon]}k^{b}k^{c}=0  ~.\label{weyl}
\end{eqnarray}
This equations implies that the KNTN geometry is of \emph{``Petrov-Pirani type D''} and these photon
trajectories are \emph{``doubly degenerate principal null congruences''}.

(b) \emph{The general case ($L \neq aE$)}.

The equations computing the radius  $r_{c}$ of the unstable circular `photon orbit' at $E=E_{c}$ and $L=L_{c}$
by introducing the impact parameter $D_{c}=\frac{L_{c}}{E_{c}}$ are
\begin{eqnarray}
r_{c}^{2}+n^2+\left(\frac{2Mr_{c}-Q^2+2n^2}{(r_{c}^2+n^2)^2}\right)\left(a-D_{c}\right)^2
+\left(a^2-D_{c}^2\right) &=& 0  ~.\label{dc}\\
r_{c}-\left[\frac{Mr_{c}^2-Q^2r_{c}-Mn^2-2n^2r_{c}}{(r_{c}^2+n^2)^2}\right]\left(a-D_{c}\right)^2 &=& 0 ~.\label{dc1}
\end{eqnarray}
From the equation (\ref{dc1}) we get
\begin{eqnarray}
D_{c}&=& a\mp \sqrt{\frac{r_{c}(r_{c}^2+n^2)^2}{Mr_{c}^2-Q^2r_{c} +2n^2r_{c}-Mn^2}} ~.\label{dc2}
\end{eqnarray}

The equation (\ref{dc}) is valid if and only if $\mid D_{c}\mid>a$. For counter rotating
orbit, we have $|D_{c}-a|=-(D_{c}-a)$, which corresponds to upper sign in the above equation and co-rotating
$|D_{c}-a|=+(D_{c}-a)$, which corresponds to lower sign in the above equation. Inserting equation (\ref{dc2})
in (\ref{dc}), we get the equation for \emph{ circular photon orbit}:
$$
r_{c}^3-3Mr_{c}^2-(3n^2-2Q^2)r_{c} \pm
$$
\begin{eqnarray}
 2a \sqrt{r_{c}[Mr_{c}^2+(2n^2-Q^2)r_{c}-Mn^2]}+Mn^2 &=& 0  ~.\label{rnul}
\end{eqnarray}
Let $r_{c}=r_{cpo}$ be the real positive  root of Eq. (\ref{rnul}). The existence condition for time-like
circular geodesics is $r>r_{cpo}$. The photon orbit with radius $r=r_{cpo}$ is the closest possible
circular orbit to the KNTN black hole.

When $Q=0$, we recover the circular photon orbit(CPO) equation of KTN space-time\cite{chur}. When $n=Q=0$,
we recover the well known  circular photon orbit of Kerr black hole \cite{bpt}. When $n=0$, we recover the circular
photon orbit(CPO) equation of KN black hole \cite{dad}.

Another important relation can be derived using equations (\ref{dc}) and (\ref{dc2}) for null circular orbits are
\begin{eqnarray}
D_{c}^2&=& a^2+(r_{c}^2+n^2) \left[\frac{3Mr_{c}^2+2(2n^2-Q^2)r_{c}-Mn^2}{Mr_{c}^2+(2n^2-Q^2)r_{c}-Mn^2} \right]~.\label{dc3}
\end{eqnarray}
Now we will derive an important physical quantity associated with the null circular geodesics is the angular
frequency measured by asymptotic observer which is denoted by $\Omega_{c}$
\begin{eqnarray}
\Omega_{c}=\frac{\left[\left(r_{c}^2-2Mr_{c}+Q^2-n^2\right)D_{c}
+\left(2Mr_{c}-Q^2 +2n^2 \right)a\right]}{\left[(r_{c}^2+
n^2)(r_{c}^2+n^2+a^2)+a^2(2Mr_{c}-Q^2+2n^2)\right]
-a\left(2Mr_{c}-Q^2+2n^2\right)D_{c}}\nonumber\\
=\frac{1}{D_{c}} ~.\label{omnul}
\end{eqnarray}
Using equations  (\ref{dc2}) and (\ref{dc}) we can see that the angular frequency $\Omega_{c}$ of the circular
null geodesics is inverse of the impact parameter $D_{c}$, which generalizes the result of KN
case\cite{sch} to the KNTN geometry. It proves that this is a \emph{general properties} of
any stationary space-time.

\section{\label{cto} The Circular Time-like Geodesics:}

For circular time-like geodesics, the equation (\ref{radial}) may be written as by
putting $\epsilon=-1$:
$$
(r^2+n^2)^2\dot{r}^{2} = E^2(r^2+n^2)^2+\left(2Mr-Q^2+2n^2\right)\left(aE-L\right)^2
$$
\begin{eqnarray}
+\left(a^2E^2-L^2\right)(r^2+n^2)- \Delta (r^2+n^2)   ~.\label{rtime}
\end{eqnarray}
where $E$ is  now to be interpreted as the  energy per unit mass of the particle describing  the trajectory.
Here the dot denote derivative with respect to the proper time $\tau$.

(a)\emph{ The special case, $L=aE$.}

Time-like geodesics with $L=aE$, like the null geodesics with $D_{c}=a$, are of some interest in that their
behaviour as they cross the the horizons are characteristic of the orbits in general. When $L=aE$,
equation (\ref{rtime}) becomes
\begin{eqnarray}
(r^2+n^2)^2\dot{r}^{2} &=& r^2(E^2-1)+2Mr-(a^2+Q^2)+(E^2+1)n^2,   ~\label{rtime1}
\end{eqnarray}
while the equations for $\dot{\phi}$ and $\dot{t}$ are the same as for the null geodesics:

\begin{eqnarray}
\dot{\phi} &=& \frac{d\phi}{d\tau}=  \frac{aE}{\Delta} ~.\label{uphir1}\\
\dot{t} &=& \frac{dt}{d\tau} =\frac{(r^2+n^2+a^2)E}{\Delta} ~.\label{utr1}
\end{eqnarray}
Eq. (\ref{rtime1}) on integration gives:
\begin{eqnarray}
\tau  &=& \int  \frac{\sqrt{r^2+n^2} dr}{\sqrt{r^2(E^2-1)+2Mr-(a^2+Q^2)+(E^2+1)n^2}},   ~\label{rtime2}
\end{eqnarray}
(b) \emph{The Circular and associated orbits:}

Now we shall find the radial equation of ISCO  which governing the time-like
circular geodesics in terms of reciprocal radius $u=1/r$ as the independent
variable, may be expressed as
$$
{\cal F}(u) = (1+n^2u^2)^2 u^{-4}\dot{u}^2=n^4u^4E^2+(2n^2-Q^2)u^4\left(aE-L\right)^2-n^2(a^2+Q^2-n^2)u^4
$$
$$
+n^2 \left(a^2E^2-L^2\right)u^4+2M(L-aE)^2u^3+2Mn^2u^3+2n^2E^2u^2+ \left(a^2E^2-L^2\right)u^2
$$
\begin{eqnarray}
-(a^2+Q^2-n^2)u^2-n^2u^2+2Mu-1+E^2 ~.\label{vu}
\end{eqnarray}
The conditions for the occurrence of circular orbit are at $r=r_{0}$ or the reciprocal radius
at $u=u_{0}$:
\begin{eqnarray}
{\cal F}(u) &=& 0 ~.\label{vu0}
\end{eqnarray}
and
\begin{eqnarray}
{\cal F}'(u) &=& 0 ~.\label{dvdu}
\end{eqnarray}
Now setting $x=L_{0}-aE_{0}$, where $L_{0}$ and $E_{0}$ are the values of energy and angular momentum for
circular orbits at the radius $r_{0}=\frac{1}{u_{0}}$.
Therefore using (\ref{vu},~\ref{dvdu}) we get the following equations
$$
E_{0}^2 (1+n^2u_{0}^2)^2 + (2n^2-Q^2)x^2u_{0}^4-n^2(a^2+Q^2-n^2)u_{0}^4
$$
$$
-(x^2+2axE_{0})(1+n^2u_{0}^2)u_{0}^2 +2Mx^2u_{0}^3+2n^2Mu^3
$$
\begin{eqnarray}
-(a^2+Q^2)u_{0}^2+2Mu_{0}-1 &=& 0  ~.\label{x1}
\end{eqnarray}
and
$$
2n^2u_{0}(1+n^2u_{0}^2)E_{0}^2 + 2(2n^2-Q^2)x^2u_{0}^3 -2n^2(a^2+Q^2-n^2)u_{0}^3
$$
$$
-(x^2+2axE_{0})(1+2n^2u_{0}^2)u_{0} +3Mx^2u_{0}^2
$$
\begin{eqnarray}
+3n^2Mu_{0}^2-(a^2+Q^2)u_{0}+M &=& 0  ~.\label{x2}
\end{eqnarray}

Equations (\ref{x1}) and (\ref{x2}) can be combined to give
$$
E_{0}^2(1+n^2u_{0}^2)^2 = (1+n^2u_{0}^2)^2-Mu_{0}(1+n^2u_{0}^2)^2
$$
\begin{eqnarray}
+Mx^2(1-n^2u_{0}^2) u_{0}^3+x^2(2n^2-Q^2)u_{0}^4  ~.\label{e2}
\end{eqnarray}
and
$$
2axE_{0}u_{0}(1+n^2u_{0}^2) = x^2 [Mu_{0}(3-n^2u_{0}^2)+3n^2u_{0}^2-2Q^2u_{0}^2-1]u_{0}
$$
\begin{eqnarray}
-(1+n^2u_{0}^2)[(a^2+Q^2-2n^2)u_{0}-M(1-n^2u_{0}^2)] ~.\label{axe}
\end{eqnarray}

By eliminating $E_{0}$ between these equations, we get the following
quadratic equation for $x^2$ i.e.,
\begin{eqnarray}
{\cal A}x^4+{\cal B}x^{2}+{\cal C} &=& 0   ~.\label{quad}
\end{eqnarray}
where we have defined
$$
{\cal A} = u_{0}^2[\{Mu_{0}(3-n^2u_{0}^2)+3n^2u_{0}^2-2Q^2u_{0}^2-1\}^2-
$$
\begin{eqnarray}
4a^2u_{0}^3\{M(1-n^2u_{0}^2)+(2n^2-Q^2)u_{0}\}] ~.\label{aktn}
\end{eqnarray}
$$
{\cal B} = -2u_{0} (1+n^2u_{0}^2)[\{Mu_{0}(3-n^2u_{0}^2)+3n^2u_{0}^2-2Q^2u_{0}^2-1\}\{(a^2+Q^2-2n^2)u_{0}
$$
\begin{eqnarray}
-M(1-n^2u_{0}^2)\} +2ua^2(1+n^2u_{0}^2)(1-Mu_{0})]
\end{eqnarray}
\begin{eqnarray}
{\cal C} &=& (1+n^2u_{0}^2)^2 \left[(a^2+Q^2-2n^2)u_{0}-M(1-n^2u_{0}^2)\right]^2
\end{eqnarray}
The solution of the equation (\ref{quad}) is
\begin{eqnarray}
x^2 &=& \frac{-{\cal B} \pm {\cal D}}{2 {\cal A}}   ~.\label{root}
\end{eqnarray}
where the discriminant of this equation is
\begin{eqnarray}
{\cal D} &=& 4au_{0} (1+n^2u_{0}^2) \, \Delta_{u_{0}} \sqrt{Mu_{0}(1-n^2u_{0}^2)+2n^2u_{0}^2-Q^2u_{0}^2} ~.\label{discri}
\end{eqnarray}
and
\begin{eqnarray}
\Delta_{u_{0}} &=& (a^2+Q^2-n^2)u_{0}^2-2Mu_{0}+1   ~.\label{deltau}
\end{eqnarray}
The solution of Eq. (\ref{aktn}) takes a particularly simpler form  by writing
$$
[Mu_{0}(3-n^2u_{0}^2)+3n^2u_{0}^2-2Q^2u_{0}^2-1]^2-
$$
\begin{eqnarray}
4a^2u_{0}^2[Mu_{0}(1-n^2u_{0}^2)+(2n^2-Q^2)u_{0}^2] = Z_{+}\,Z_{-} ~.\label{z+}
\end{eqnarray}
where
$$
Z_{\pm} = 1-Mu_{0}(3-n^2u_{0}^2)+2Q^2u_{0}^2-3n^2u_{0}^2
$$
\begin{eqnarray}
\pm 2au_{0}\sqrt{Mu_{0}(1-n^2u_{0}^2)+(2n^2-Q^2)u_{0}^2} ~.\label{z+-}
\end{eqnarray}
Thus we get the solution as
\begin{eqnarray}
x^2u_{0}^2 &=& \frac{-{\cal B} \pm {\cal D}}{Z_{+} Z_{-}}   ~.\label{x2u2}
\end{eqnarray}
Thus, we find
\begin{eqnarray}
x^2u_{0}^2 &=& (1+n^2u_{0}^2)\frac{\Delta_{u}-Z_{\mp}}{Z_{\mp}}   ~.\label{zz}
\end{eqnarray}
On the other hand (as we may verify),
\begin{eqnarray}
\Delta_{u_{0}}-Z_{\mp} &=& u_{0} \left[a\sqrt{u_{0}} \pm \sqrt{M(1-n^2u_{0}^2)+(2n^2-Q^2)u_{0}} \right]^2  ~.\label{za}
\end{eqnarray}
Therefore the solution for $x$ thus may be written as
\begin{eqnarray}
x &=& - \sqrt{1+n^2u_{0}^2}\frac{\left[a\sqrt{u_{0}} \pm \sqrt{M(1-n^2u_{0}^2)+(2n^2-Q^2)u_{0}}\right]}{\sqrt{u_{0}{Z}_{\pm}}}
~.\label{solx}
\end{eqnarray}
It will appear here that the upper sign in the foregoing equations applies to counter-rotating orbits, while the lower sign
applies to co-rotating orbits. This convention will be adhered in the subsequent analysis in this section.

Replacing the solution (\ref{solx}) for $x$ in equation (\ref{e2}), we get
the energy for circular orbit:
$$
E_{0}= \frac{1}{\sqrt{(1+n^2u_{0}^2){Z}_{\mp}}} \times
$$
\begin{eqnarray}
\left[1-2Mu_{0}+Q^2u_{0}^2-n^2u_{0}^2\mp au_{0}\sqrt{Mu_{0}(1-n^2u_{0}^2)-(2n^2-Q^2)u_{0}^2} \right] ~.\label{eng} \nonumber\\
\end{eqnarray}
and the value of angular momentum to be associated with this value of $E_{0}$
along the circular orbit is:
$$
L_{0} = aE_{0}+x
$$
or
$$
L_{0} = \mp \frac{1}{\sqrt{u_{0}{Z}_{\mp}(1+n^2u_{0}^2)}} \times
$$
\begin{eqnarray}
\left[\left(1+a^2u_{0}^2+n^2u_{0}^2 \right)\sqrt{M(1-n^2u_{0}^2)+u_{0}(2n^2-Q^2)}
\pm 2aM\sqrt{u_{0}^3}\pm a(2n^2-Q^2)\sqrt{u_{0}^5}\right] ~.\label{ang} \nonumber\\
\end{eqnarray}
As we previously defined $E_{0}$ and $L_{0}$ followed by equations (\ref{eng}) and (\ref{ang}) are the
energy and the angular momentum per unit mass of a particle
describing a circular orbit of radius $u_{0}$.

Therefore the minimum radius for a stable circular orbit will be obtained at a point of inflection of
the function ${\cal F}(u)$ i.e. we have to supply equations (\ref{vu0}, ~\ref{dvdu}) with the further
equation
\begin{eqnarray}
{\cal F}''(u) &=& 0 ~.\label{d2vdu2}
\end{eqnarray}
After some long algebraic computation, one may obtain the ISCO equation for KNTN space-time:
$$
M-6M^2u_{0}+(9MQ^2-3Ma^2-15Mn^2)u_{0}^2+(4a^2Q^2-4Q^4+4M^{2}n^2+16n^2Q^2-8a^2n^2)u_{0}^3
$$
$$
+(15Mn^4-6Mn^2Q^2+6Mn^2a^2)u_{0}^4-6M^2n^4u_{0}^5+(Mn^4Q^2-Mn^6+Ma^2n^4)u_{0}^6
$$
\begin{eqnarray}
\mp 8a[Mu_{0}(1-n^2u_{0}^2)+(2n^2-Q^2)u_{0}^2]^{3/2} &=& 0~.\label{isco}
\end{eqnarray}
Reverting to the variable $r_{0}$, we obtain the equation of ISCO for non-extremal KNTN
black-hole  reads as:
$$
Mr_{0}^6-6M^2r_{0}^5+(9MQ^2-3Ma^2-15Mn^2)r_{0}^4 +(4a^2Q^2-4Q^4+4M^{2}n^2+16n^2Q^2-8a^2n^2)r_{0}^3
$$
$$
+(15Mn^4-6Mn^2Q^2+6Mn^2a^2) r_{0}^2-6M^2n^{4}r_{0} \mp 8a [Mr_{0}(r_{0}^2-n^2)+r_{0}^{2}(2n^2-Q^2)]^{3/2}
$$
\begin{eqnarray}
+(Mn^4Q^2+Ma^2n^4-Mn^6) &=& 0 ~.\label{isco2}
\end{eqnarray}
Let $r_{0}=r_{ISCO}$  be the smallest real root of the equation, which will be the ISCO
of the black-hole. Here $(-)$ sign is  for  the direct orbit $(L>0)$ and $(+)$ sign  is
for  the retrograde orbit $(L<0)$. It may be noted that when $Q=0$, we recover the ISCO equation for
KTN spacetime \cite{chur}. When $n=0$, we recover the ISCO equation for KN
black hole\cite{dad}. Finally when $Q=n=0$, we recover the ISCO equation for
Kerr black hole \cite{bpt}.

(c) \emph{Effective Potential:}

The stability properties of the circular geodesics around the KNTN space-time
can be determined by using the effective potential method. Thus one can write
the effective potential for massive particles that governing the radial motion
is given by
\begin{eqnarray}
\frac{E^2-1}{2}&=& \frac{1}{2}\left(\frac{dr}{d\tau}\right)^{2}+{\cal V}_{eff}
~. \label{veff}
\end{eqnarray}
where the effective potential is given by

\begin{eqnarray}
{\cal V}_{eff} &=& \frac{(r^{2}+n^{2})[2aEx+\Delta-(r^{2}+n^{2})]+x^{2}(\Delta-a^2)}{2(r^{2}+n^{2})^{2}}
~. \label{veffkn}
\end{eqnarray}
First we recall the radial time-like geodesics $(L=aE)$ for which the effective potential
reduces to
\begin{eqnarray}
{\cal V}_{eff} &=& \frac{a^2+Q^2-2n^2-2Mr}{2(r^{2}+n^{2})}
~. \label{rveff}
\end{eqnarray}
The behaviour of this effective potential could be seen from the following Fig. \ref{rp} and
Fig. \ref{rp1}. In this plot, we also show how the effective potential changes with $r$
for time-like radial geodesics for different values of NUT parameter.
\begin{figure}
\begin{center}
{\includegraphics[width=0.45\textwidth]{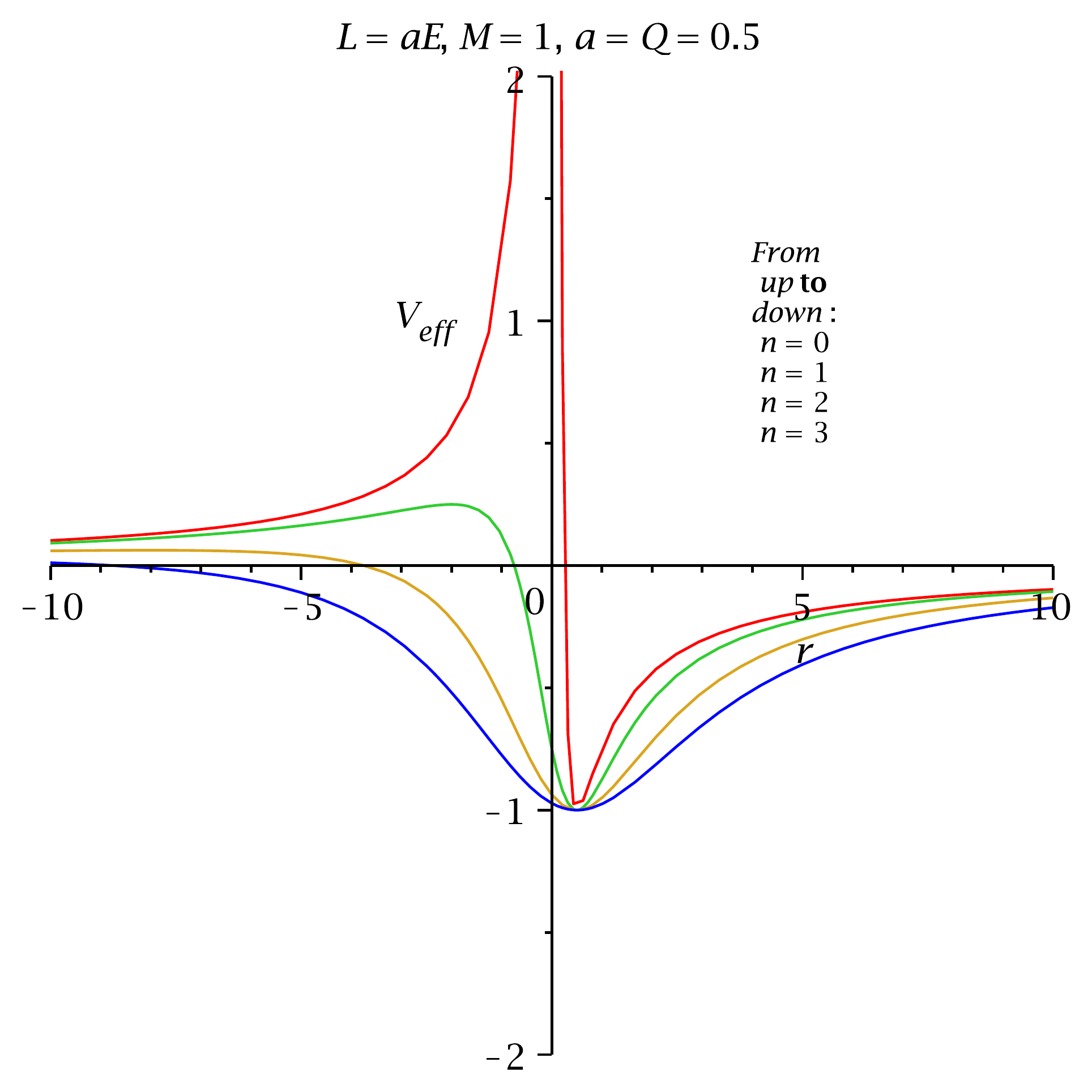}}
{\includegraphics[width=0.45\textwidth]{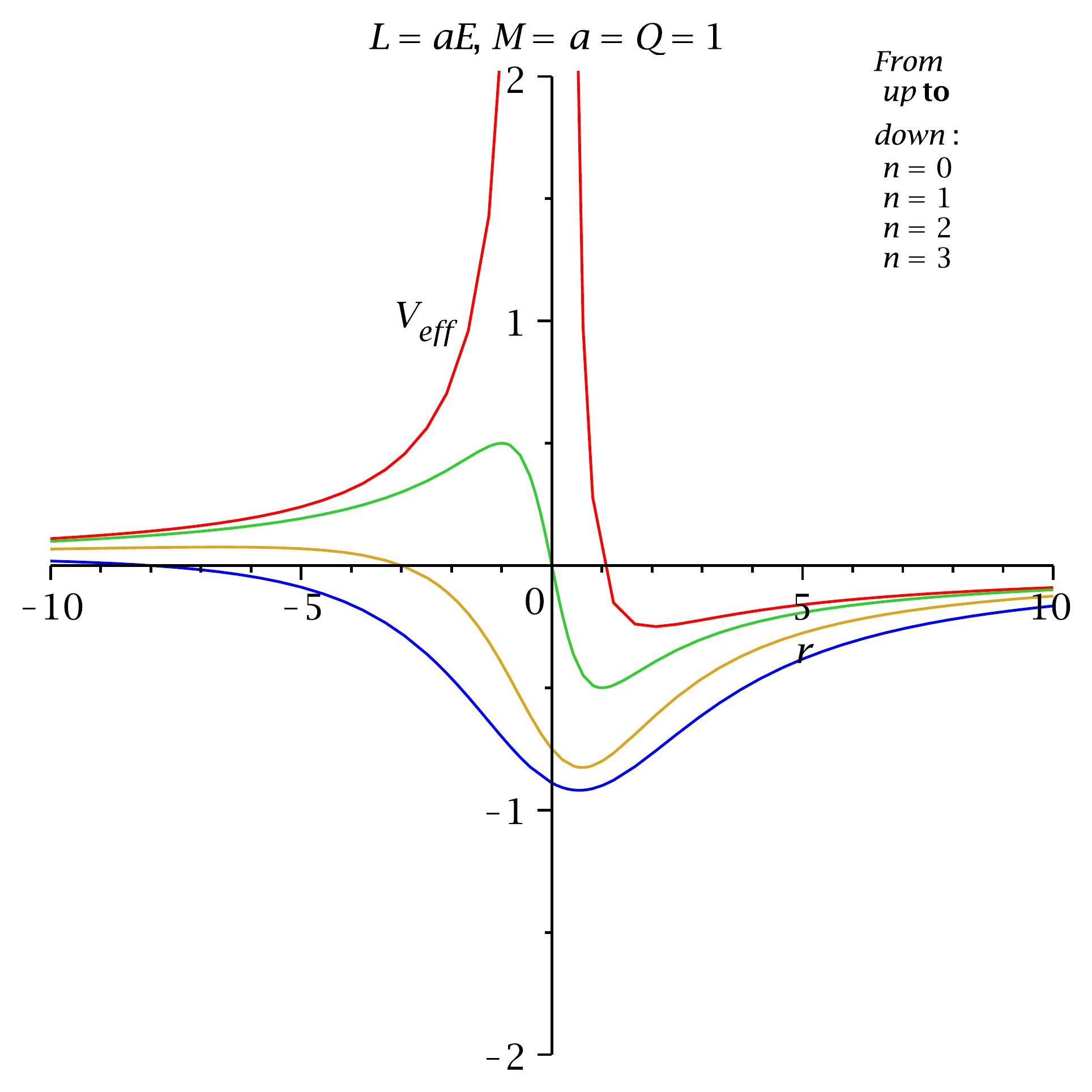}}
\end{center}
\caption{The figure shows the variation  of $V_{eff}$  with $r$ for various values of NUT parameter.
\label{rp}}
\end{figure}
\begin{figure}
\begin{center}
{\includegraphics[width=0.45\textwidth]{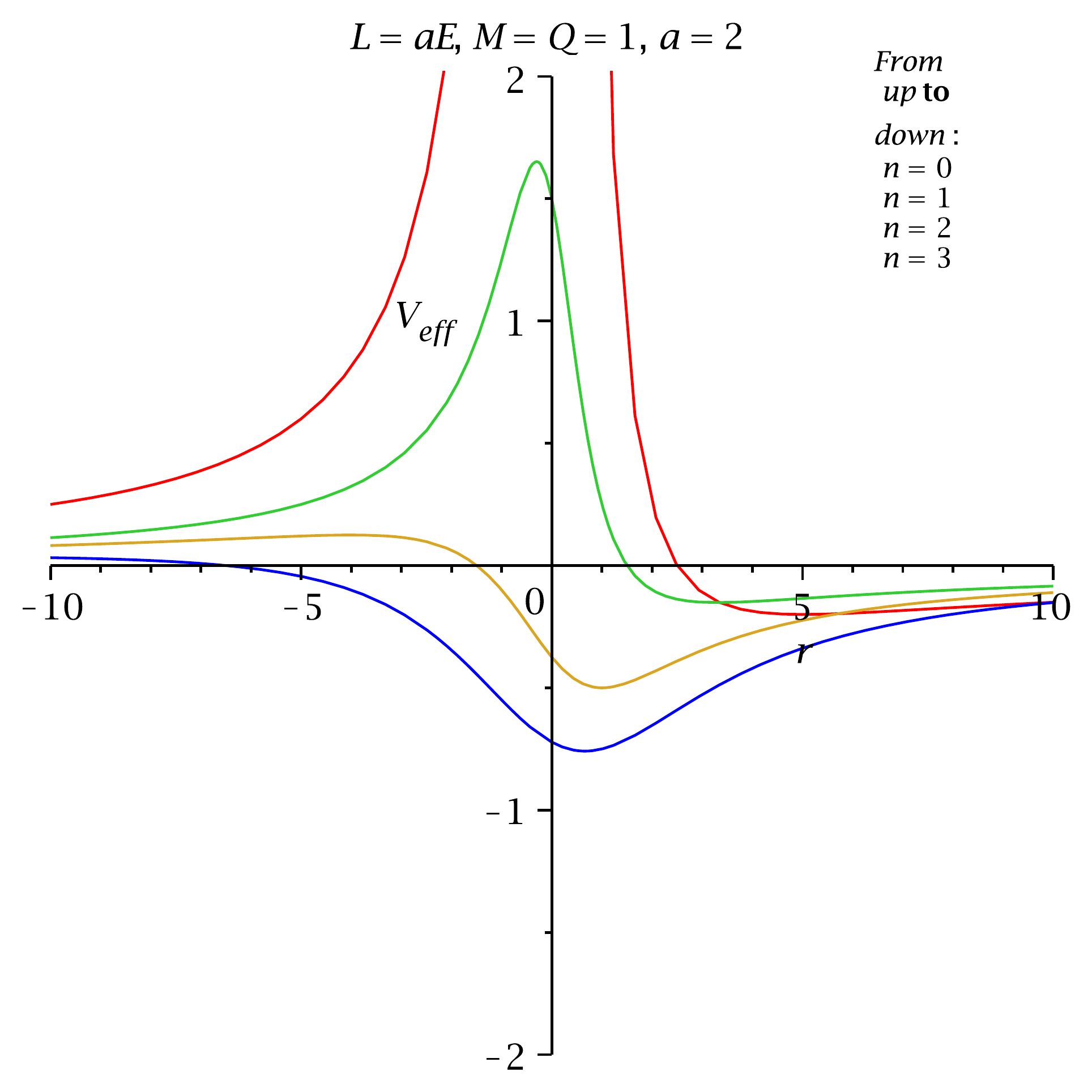}}
\end{center}
\caption{The figure shows the variation  of $V_{eff}$  with $r$ for various values of NUT parameter.
\label{rp1}}
\end{figure}

\begin{figure}
\begin{center}
{\includegraphics[width=0.45\textwidth]{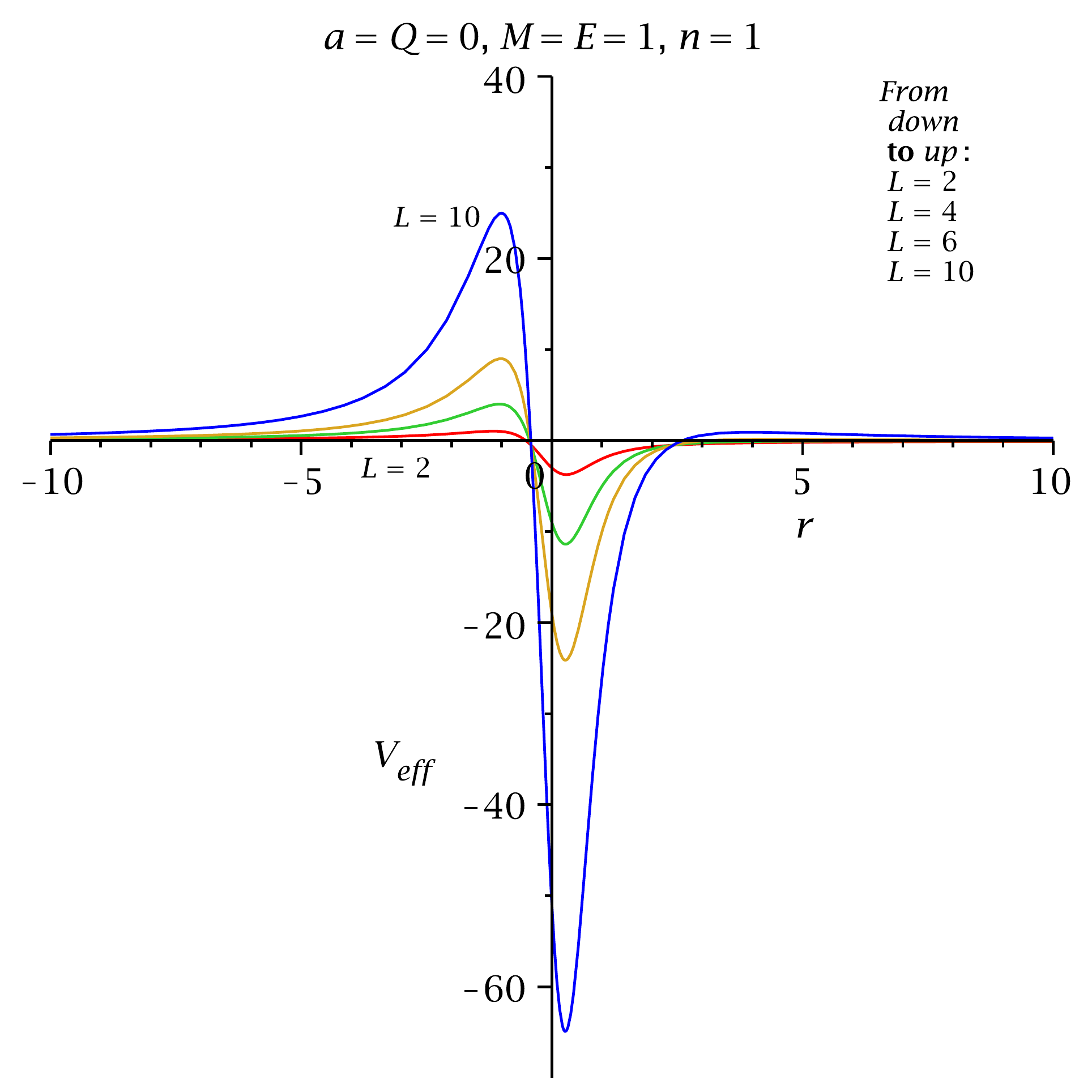}}
\end{center}
\caption{The figure shows the variation  of $V_{eff}$  with $r$ for TN black hole.
\label{ep0}}
\end{figure}

\begin{figure}
\begin{center}
{\includegraphics[width=0.45\textwidth]{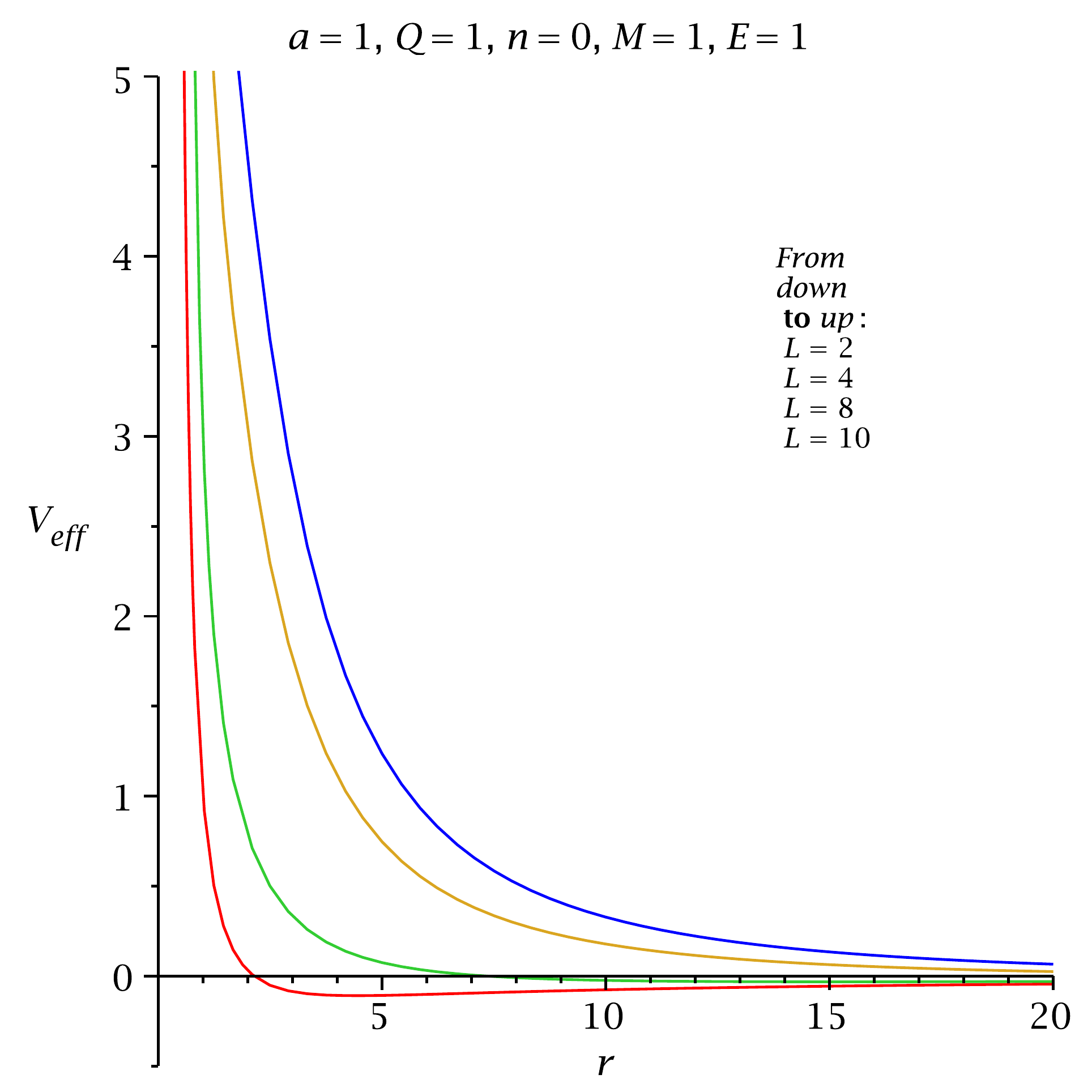}}
{\includegraphics[width=0.45\textwidth]{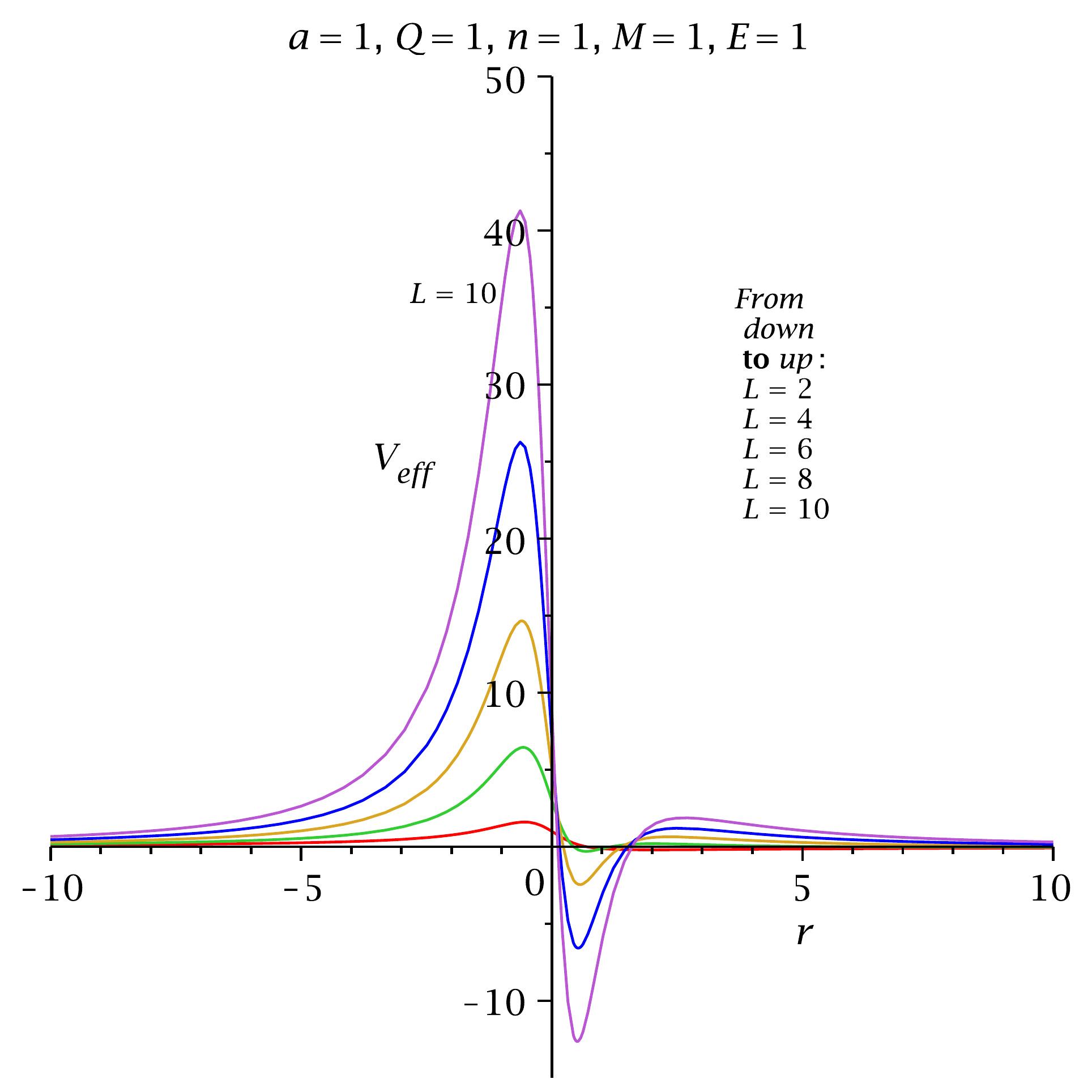}}
\end{center}
\caption{The figure shows the variation  of $V_{eff}$  with $r$ for
KN black hole and KNTN black hole.
\label{ep}}
\end{figure}

\begin{figure}
\begin{center}
{\includegraphics[width=0.45\textwidth]{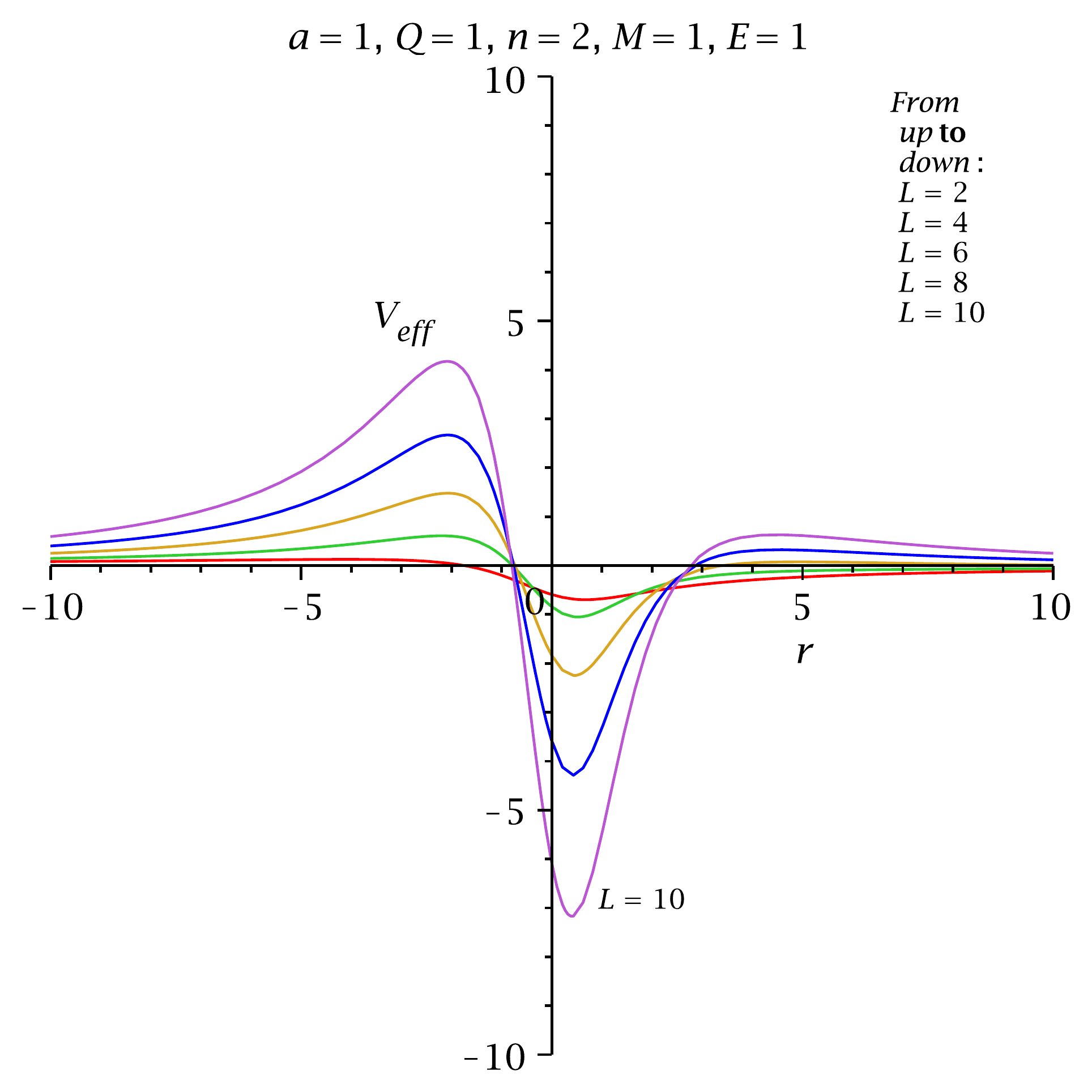}}
{\includegraphics[width=0.45\textwidth]{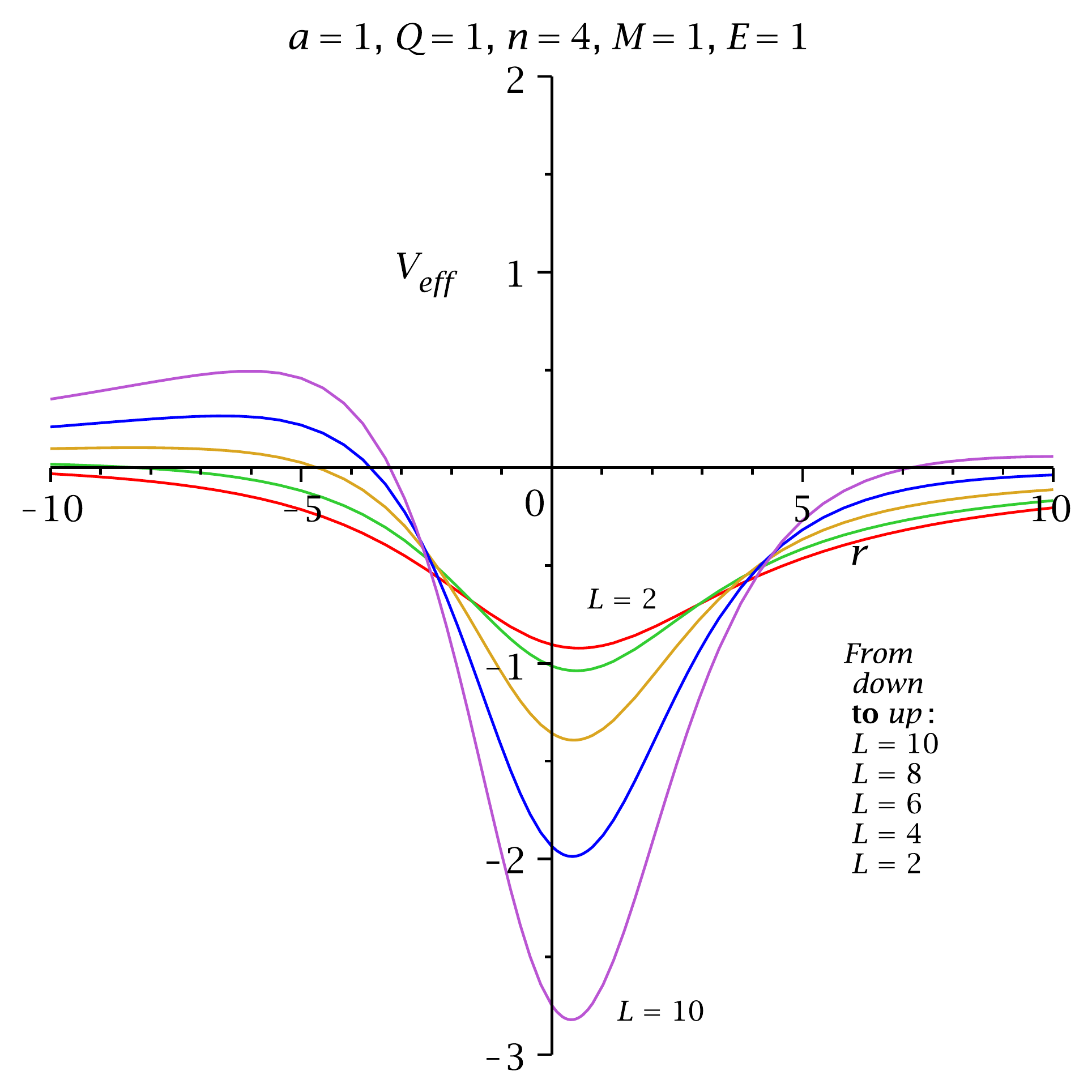}}
\end{center}
\caption{The figure shows the variation  of $V_{eff}$  with $r$ for KNTN black hole.
\label{ep1}}
\end{figure}

\begin{figure}
\begin{center}
{\includegraphics[width=0.45\textwidth]{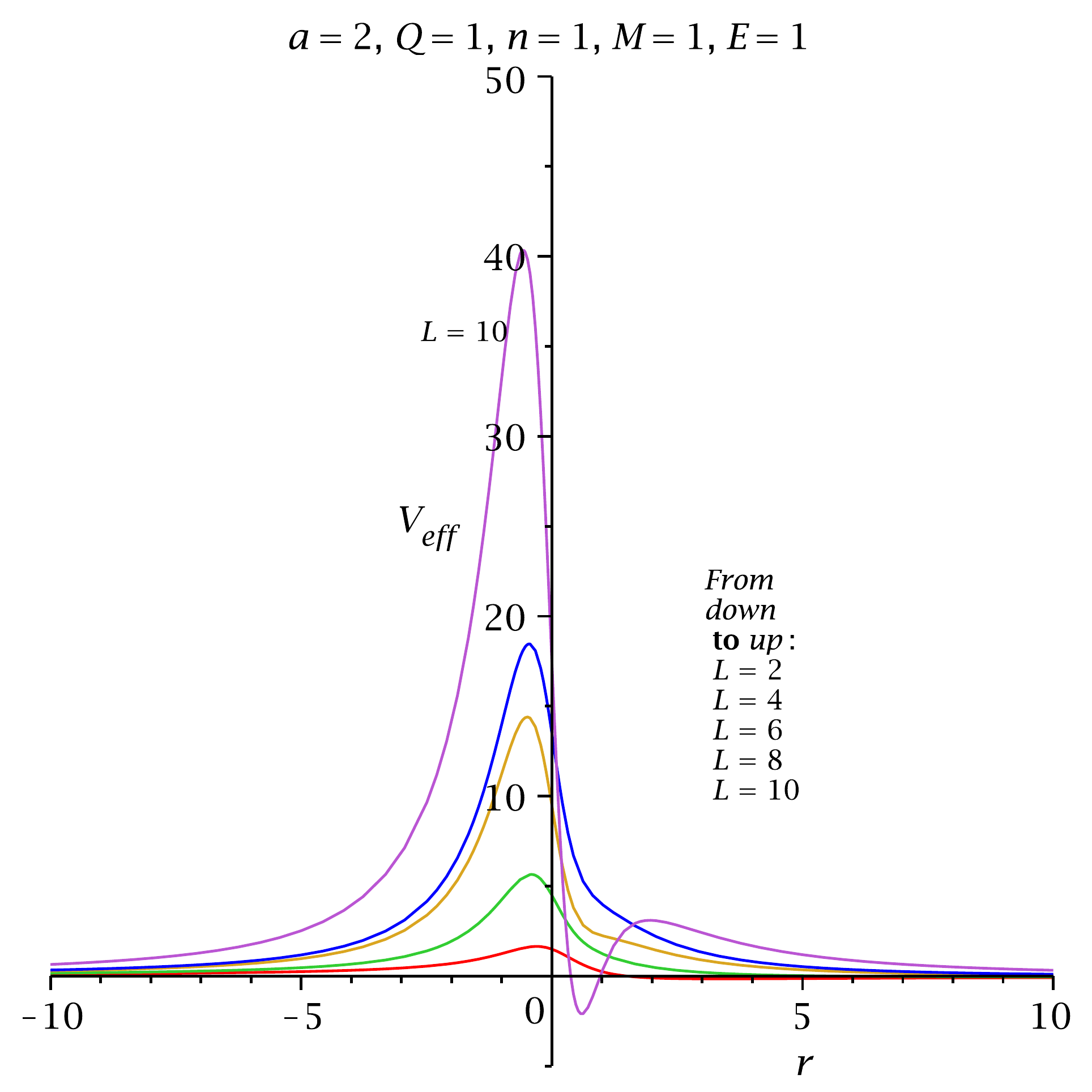}}
{\includegraphics[width=0.45\textwidth]{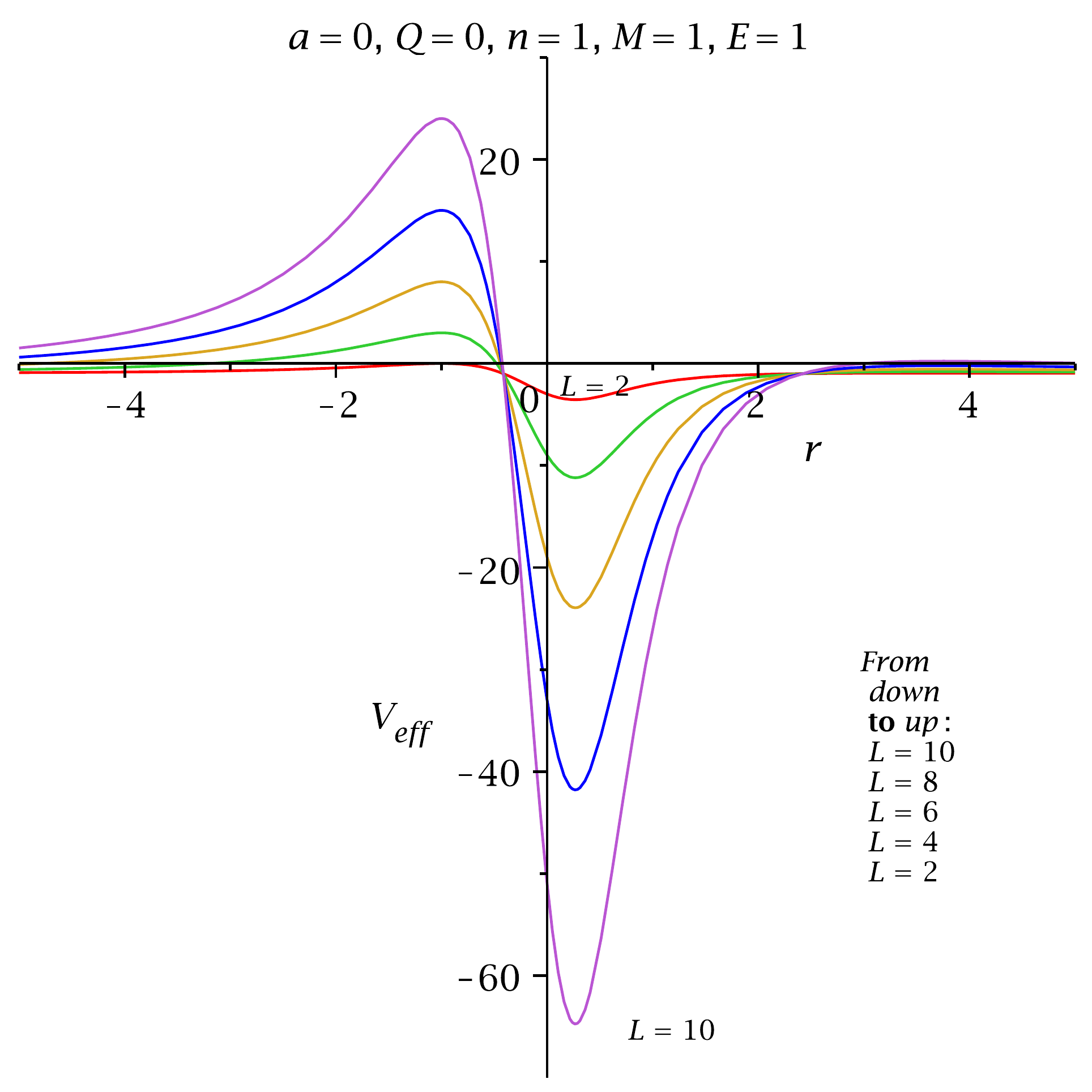}}
\end{center}
\caption{The figure shows the variation  of $V_{eff}$  with $r$ for KNTN black hole.
\label{ep2}}
\end{figure}

In Fig.\ref{ep0}, Fig.\ref{ep}, Fig.\ref{ep1} and Fig\ref{ep2}, we have plotted the
effective potential for
a massive particles for various values of angular momentum with NUT and with out NUT
parameter. The figures show the radial dependence of the effective potential with
$n=0$ and $n\neq 0$. When $n=0$, the effective potential at large radial distance
does not change much more with the increasing of angular momentum parameter. When
we incorporated the NUT parameter, the shape of the effective potential deforms in
comparison with zero NUT parameter and it also changes for different values of $L$.
Furthermore when we increase the value of NUT
parameter for a fixed value of spin parameter and charge parameter, the height of
the potential barrier decreases (See Figs. \ref{ep}b and \ref{ep1}).

It may be noted that this effective potential is depends on energy and angular
momentum. Whereas for any spherically symmetric black holes, like Schwarzschild
black holes this is indeed not so. Where the effective potential is only the function
of $r$. This is an important differences between spherically symmetric effective
potential and axi-symmetric effective potential. This important differences also
reflect particularly in the Lense-Thirring (frame dragging effect) effect of the
spinning black hole.

To show the stability properties of the geodesics one may compute the second
order derivative of the effective potential. Thus the condition for the stable
circular geodesics in the equatorial plane when the effective potential must
have a minimum value. That means
\begin{eqnarray}
\frac{\partial^{2}{\cal V}_{eff}}{\partial r^{2}}|_{r=r_{0}} &>& 0
~. \label{stab}
\end{eqnarray}
with additional requirement is that:

For a particle to describe a circular orbit at constant $r=r_{0}$, the initial
radial velocity must vanish i.e. $u^{r}=\frac{dr}{d\tau}=0$. From Eq. (\ref{veff})
\begin{eqnarray}
\frac{E^2-1}{2}&=& {\cal V}_{eff}
~. \label{stab1}
\end{eqnarray}
But to stay on a circular orbit the radial acceleration must also vanish.
Differentiating Eq. (\ref{veff}) with respect to proper time $\tau$ leads
to the condition
\begin{eqnarray}
\frac{\partial{\cal V}_{eff}}{\partial r}|_{r=r_{0}} &=& 0
~. \label{stab2}
\end{eqnarray}
These three conditions evaluate the stability properties of the KNTN
black hole. Alternatively the ISCO equation (\ref{isco2}) can be obtained
at the point of inflection of the  effective potential i.e.
\begin{eqnarray}
\frac{\partial^{2}{\cal V}_{eff}}{\partial r^{2}}|_{r=r_{0}} &=& 0
~. \label{stab3}
\end{eqnarray}
with additional two equations (\ref{stab1}) and (\ref{stab2}). It is shown
that the equatorial time like circular geodesics of KNTN space-times are
\emph{stable} by computing the second order derivative of the effective potential.

(d) \emph{Effective Potential for Photon:}

The effective potential for photon could be derived by using Eq. (\ref{nul}) as
\begin{eqnarray}
\frac{E^2-1}{2}&=& \frac{1}{2}\left(\frac{dr}{d\lambda}\right)^{2}+{\cal U}_{eff}
~. \label{ueff}
\end{eqnarray}
where the effective potential is given by
\begin{eqnarray}
{\cal U}_{eff} &=& \frac{(r^{2}+n^{2})[2aEx-(r^{2}+n^{2})]+x^{2}(\Delta-a^2)}{2(r^{2}+n^{2})^{2}}
~. \label{ueffkn}
\end{eqnarray}

\begin{figure}
\begin{center}
{\includegraphics[width=0.45\textwidth]{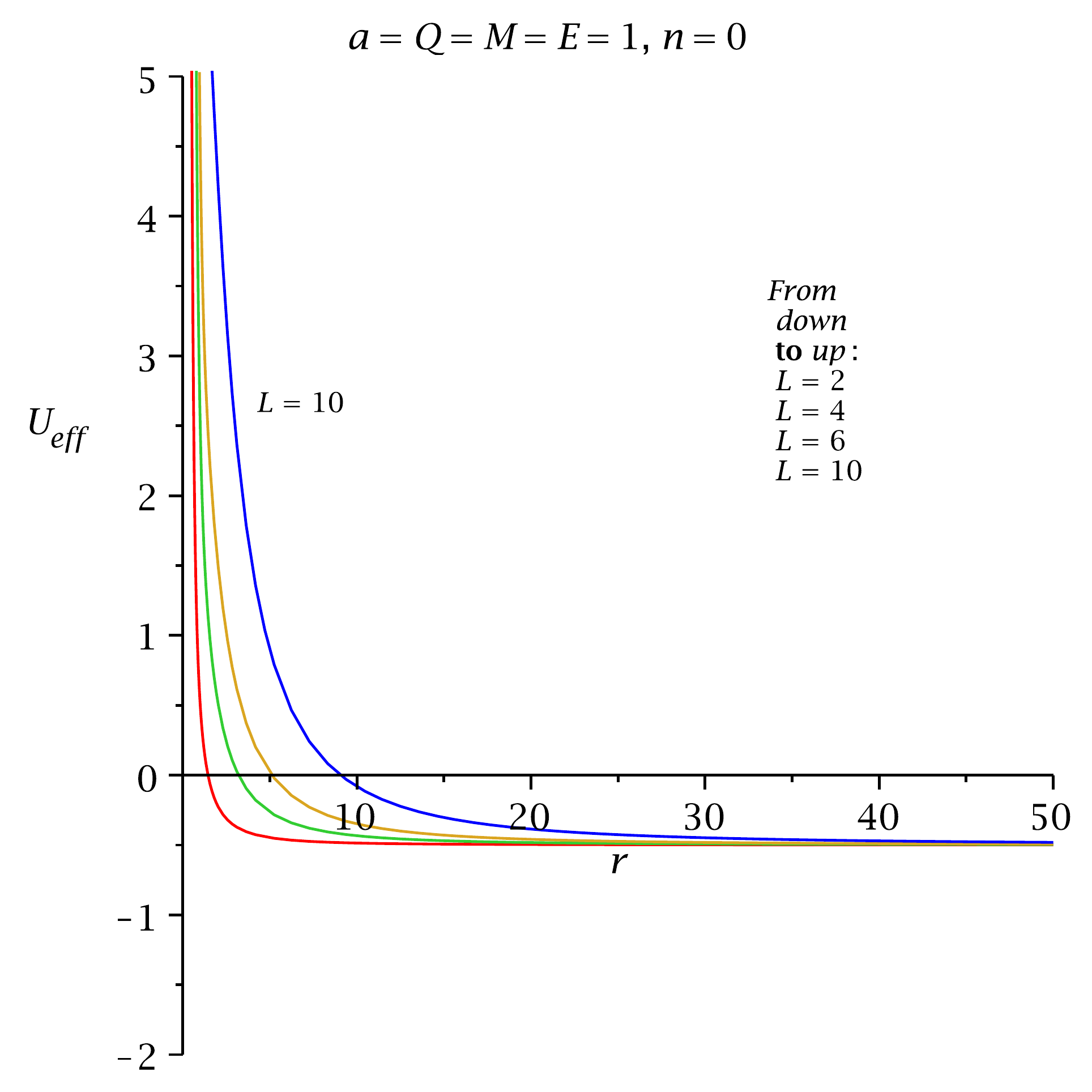}}
{\includegraphics[width=0.45\textwidth]{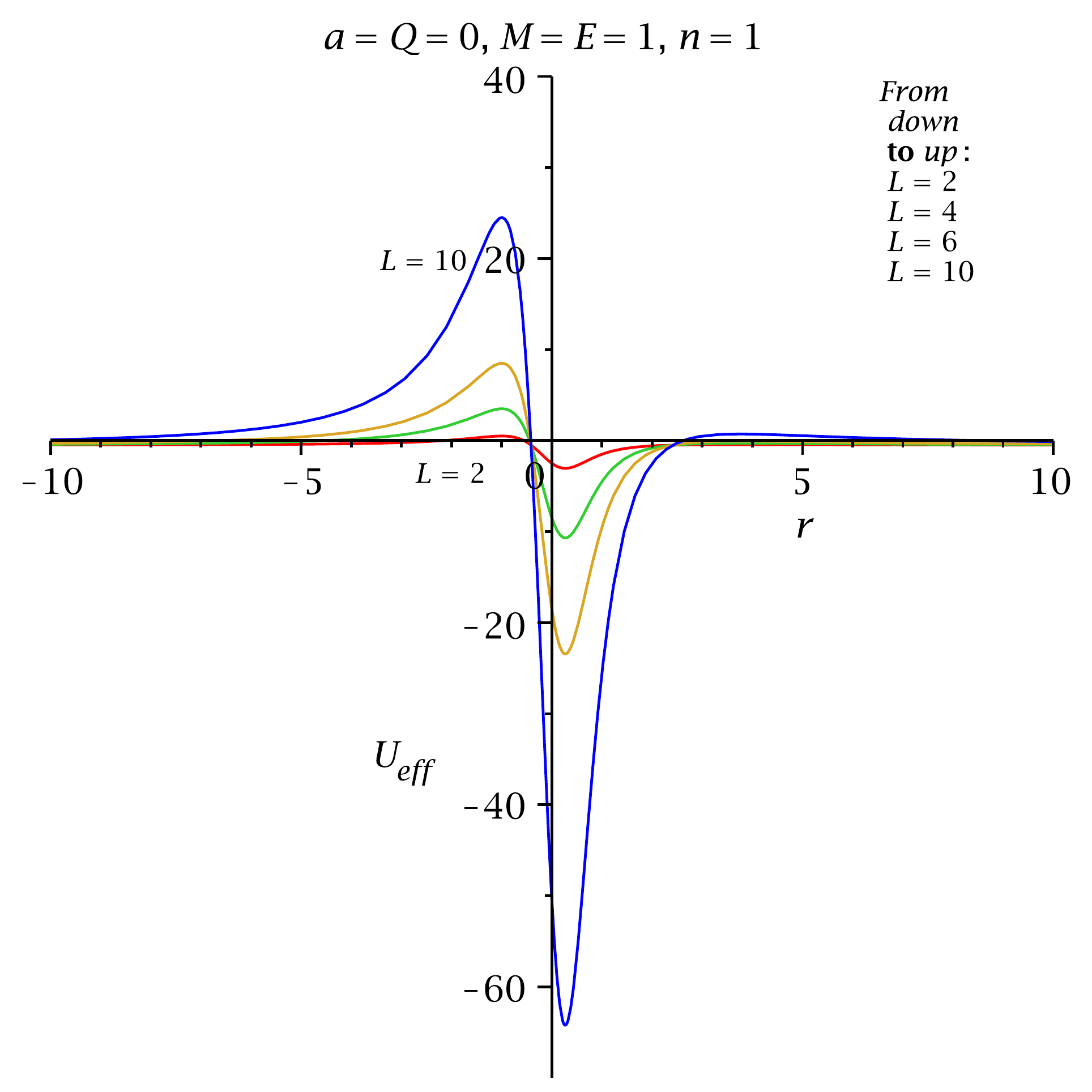}}
\end{center}
\caption{The figure shows the variation  of $U_{eff}$  with $r$ for KN black hole and TN black hole.
\label{nu1}}
\end{figure}

\begin{figure}
\begin{center}
{\includegraphics[width=0.45\textwidth]{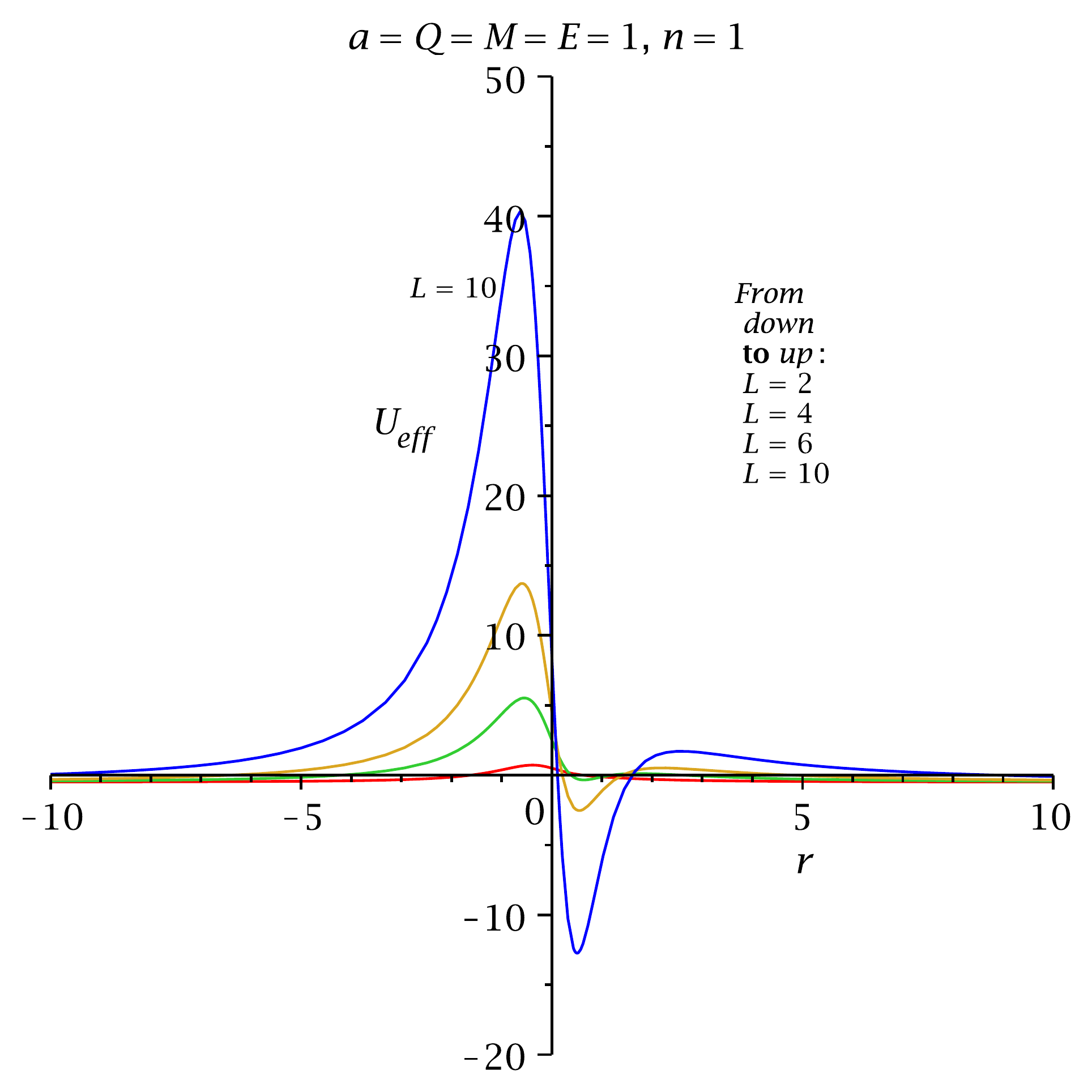}}
{\includegraphics[width=0.45\textwidth]{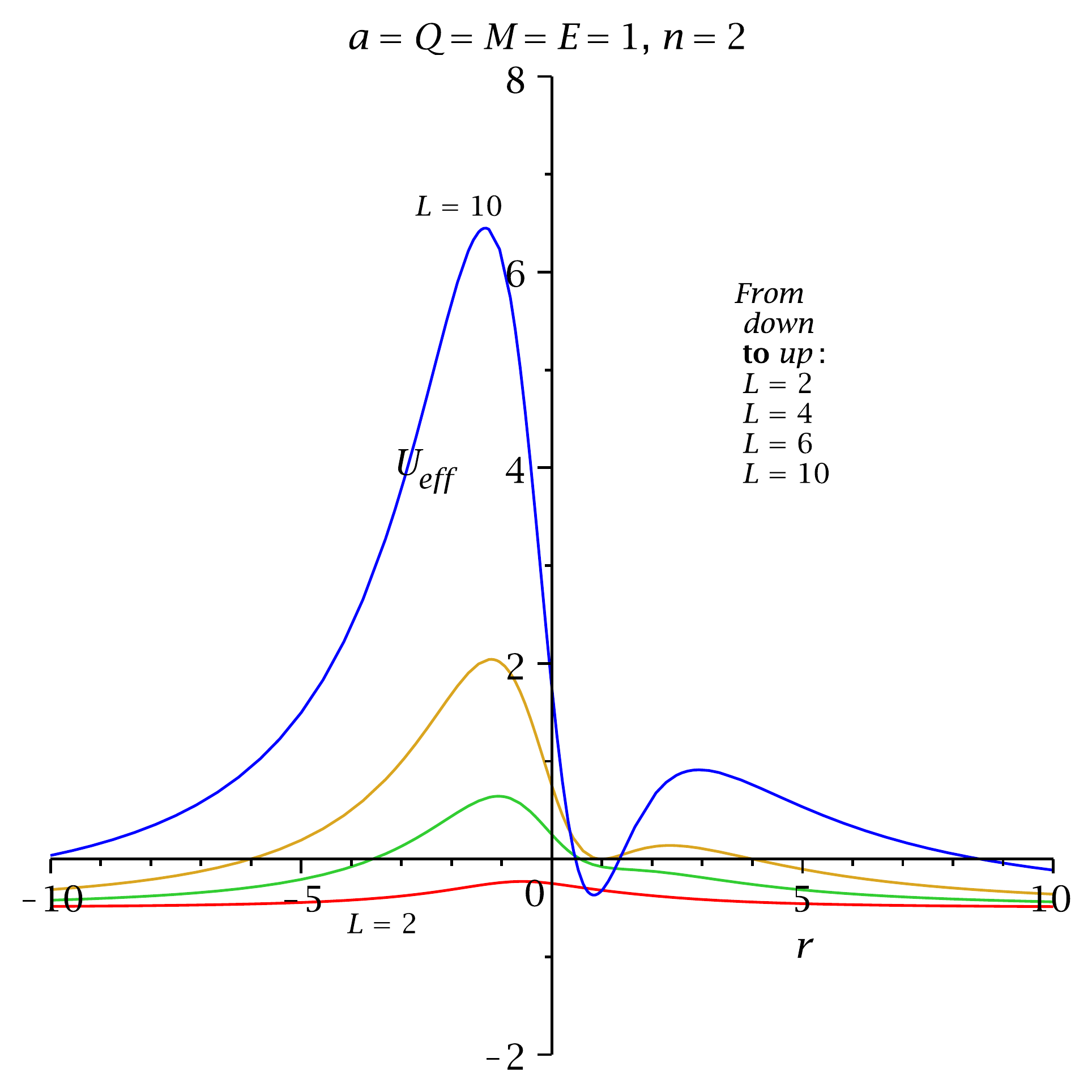}}
\end{center}
\caption{The figure shows the variation  of $U_{eff}$  with $r$ for KNTN black hole
for different values of $n$.
\label{nu2}}
\end{figure}

\begin{figure}
\begin{center}
{\includegraphics[width=0.45\textwidth]{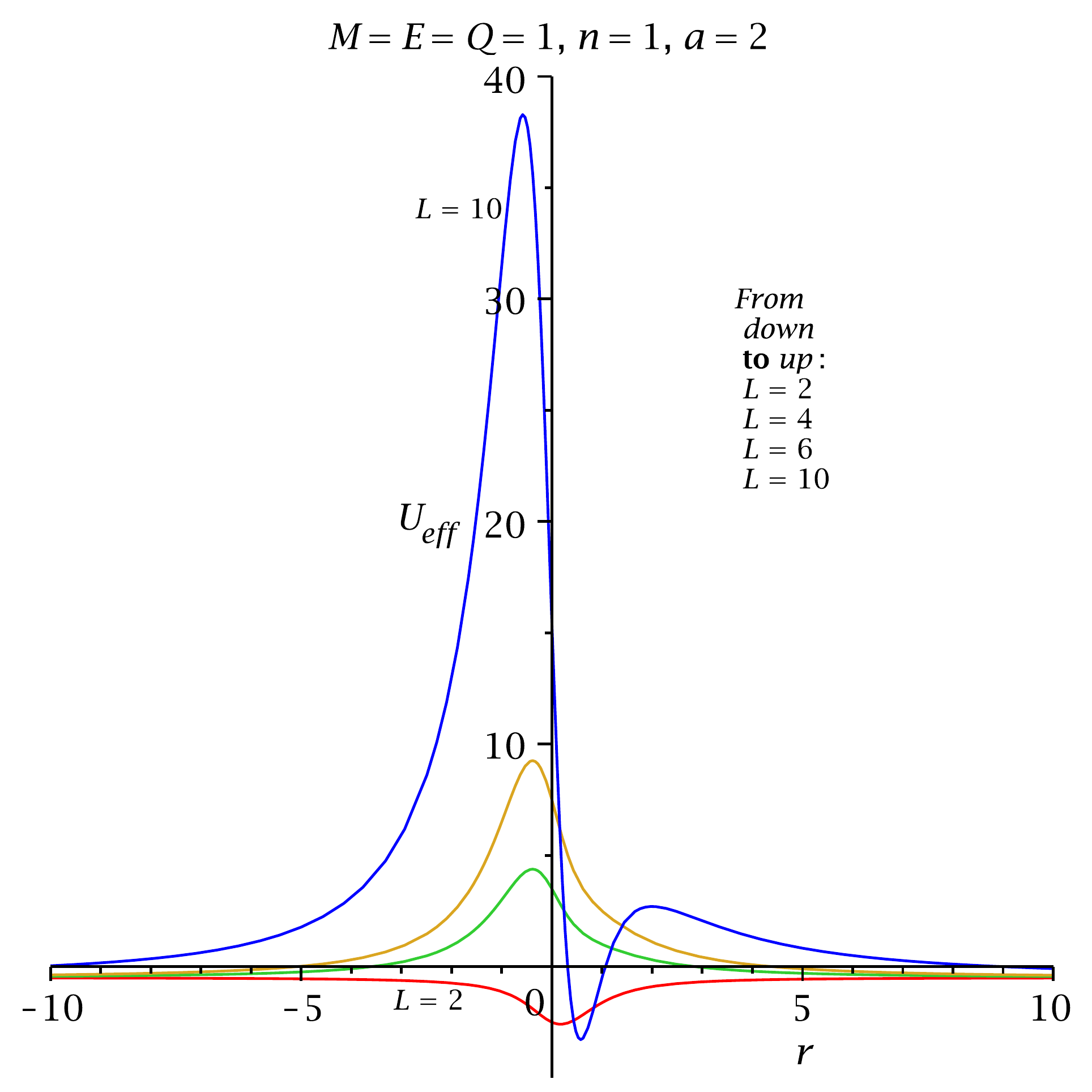}}
\end{center}
\caption{The figure shows the variation  of $U_{eff}$  with $r$ for KNTN black hole.
\label{nu3}}
\end{figure}

In Fig \ref{nu1}, Fig \ref{nu2} and Fig \ref{nu3},  we show how the
effective potential for photon changes for different values of angular
momentum and NUT parameter.


\section{\label{avtlo} Angular Velocity of Time-like Circular Orbit}

Now we compute the orbital angular velocity for time-like circular geodesics at $r=r_{0}$
is given by
\begin{eqnarray}
\Omega_{0}=\frac{\left[L_{0}(1+n^2u_{0}^2)-2Mu_{0}x-2n^2u_{0}^2x+Q^2u_{0}^2x \right]
u_{0}^2}{(1+n^2u_{0}^2)(1+a^2u_{0}^2+n^2u_{0}^2)E_{0}-au_{0}^2
(2Mu_{0}+2n^2u_{0}^2-Q^2u_{0}^2)x}~.\label{omeu}
\end{eqnarray}

After some algebra, this can be reduced to
\begin{eqnarray}
\Omega_{0}=\mp \frac{\sqrt{Mu_{0}^3 (1-n^2u_{0}^2)+(2n^2-Q^2)u_{0}^2}}{1+n^2u_{0}^2\mp
au_{0}\sqrt{Mu_{0}(1-n^2u_{0}^2)+(2n^2-Q^2)u_{0}^2}}~.\label{omef}
\end{eqnarray}
Reverting to the variable $r_{0}$, we obtain the angular velocity for time like circular orbit is
\begin{eqnarray}
\Omega_{0}=\mp \frac{\sqrt{M(r_{0}^{2}-n^2)+(2n^2-Q^2)r_{0}}}{\left[(r_{0}^2+n^2)\sqrt{r_{0}}
\mp a\sqrt{M(r_{0}^{2}-n^2)+(2n^2-Q^2)r_{0}}\right]}~.\label{omer}
\end{eqnarray}
This is also called  the Kepler frequency  in  the KNTN space-time and kepler time period could
be found by the relation $T_{0}=\frac{2\pi}{\Omega_{0}}$.

In the limit $a=Q=n=0$, this equation verifies the relativistic Kepler's law $T_{0}^{2}\propto r_{0}^{3}$ for
Schwarzschild black-hole.

The rotational velocity with respect to the locally non-rotating observers (LNRO) is given by
$$
v^{\phi}=\mp \frac{1}{\left[1+n^2u_{0}^2 \mp
au_{0}\sqrt{Mu_{0}(1-n^2u_{0}^2)+(2n^2-Q^2)u_{0}^2} \right]\sqrt{\Delta_{u_{0}}}} \times
$$
\begin{eqnarray}
\left[\left(1+a^2u_{0}^2+n^2u_{0}^2 \right)\sqrt{Mu_{0}(1-n^2u_{0}^2)+u_{0}^{2}(2n^2-Q^2)}
\pm 2aMu_{0}^2 \pm a(2n^2-Q^2)u_{0}^3 \right] ~.\label{omef1}\nonumber\\
\end{eqnarray}

We may parenthetically note here that we can recover  from Eq. (\ref{rnul}) the
condition for the occurrence of the unstable circular null geodesics by considering
limit  $E_{0}\rightarrow \infty$, when
$$
Z_{\pm} = 1-Mu_{0}(3-n^2u_{0}^2)+2Q^2u_{0}^2-3n^2u_{0}^2\pm
$$
\begin{eqnarray}
2au_{0}\sqrt{Mu_{0}(1-n^2u_{0}^2)+(2n^2-Q^2)u_{0}^2}   ~.\label{z0}
\end{eqnarray}
or alternatively
$$
r_{0}^3-3Mr_{0}^2-(3n^2-2Q^2)r_{0} \pm
$$
\begin{eqnarray}
2a \sqrt{r_{0}[Mr_{0}^2+(2n^2-Q^2)r_{0}-Mn^2]}+Mn^2 &=& 0  ~.\label{rul}
\end{eqnarray}
The above equation describes the radius of CPO (\ref{rnul}) at $r_{0}=r_{c}$. Here $(-)$
sign indicates for the direct orbit and $(+)$ sign
indicates for the retrograde orbit. The real positive root of the equation
is the closest CPO of the black-hole.

\section{\label{mbcokn} The marginally bound circular orbit:}

Besides the limiting case of null circular geodesics  $E_{0}\rightarrow \infty$, the case
of the marginally bound orbit with  $E_{0}^2=1$ is of some interest: it corresponds to
the case of a particle, at rest of infinity, falling towards the black hole.
Thus when a particle at rest at infinity falling towards the black-hole, we call the situation is
marginally bound circular orbit. Using Eqs. (\ref{eng}) and
(\ref{solx}), the radius of the marginally bound circular orbit with $E_{0}^2=1$ is given by
\begin{eqnarray}
E_{0}^2=1-Mu_{0}+\frac{\left[Mu_{0}(1-n^2u_{0}^2)+ (2n^2-Q^2)u_{0}^2\right]}{(1+n^2u_{0}^2)^{2}} x^2u_{0}^2=1 ~.\label{marg}
\end{eqnarray}
After simplification,  we obtain the radial equation for MBCO:
$$
Mr_{0}^7-4M^2r_{0}^6-Ma^2r_{0}^{5}-7Mn^2r_{0}^{5}+4MQ^2r_{0}^{5}+2M^2n^2r_{0}^{4}
$$
$$
-2n^2a^2r_{0}^{4}+a^2Q^2r_{0}^{4}-r_{0}^{2}(r_{0}^{2}+n^2)(2n^2-Q^2)^2+n^4M^2r_{0}^{2}
$$
$$
-2n^4a^2r_{0}^{2}+n^2a^2Q^2r_{0}^{2}+Mn^4a^2r_{0}+4Mn^6r_{0}-2Mn^4Q^2r_{0}
$$
$$
\mp 2ar_{0}(r_{0}^{2}+n^2)(2n^2-Q^2)\sqrt{Mr_{0}(r_{0}^{2}-n^2)+r_{0}^{2}(2n^2-Q^2)}
$$
\begin{eqnarray}
-M^2n^6 &=& 0 ~.\label{marg4}
\end{eqnarray}
Let $r_{0}=r_{mbco}$ be the real smallest root of the above equation, which
will be  the closest MBCO to the  KNTN black hole.

\section{\label{pp} The Penrose Process for KNTN Space-time:}

In this section we will show how one may extract energy from the black hole
in the presence of both NUT parameter $(n)$ and charge parameter ($Q$). The
surface on which $g_{tt}$ vanishes is called stationary limit surface. In fact,
in KNTN space-time, this surface does not coincide with the event horizon except
at the poles. In the toroidal space between the two surfaces, i.e., in the
ergo-sphere, the Killing vector $\partial_{t}$ becomes space-like and likewise,
the conserved component, $p_{t}$, of the four momentum. Thus the energy of a
particle in this finite region of space between the event horizon and the
stationary limit surface, as perceived by an observer at infinity,
can be negative.
This negative energy inside the ergo-sphere  has an important consequences
in the black hole physics. It may allows the process, what we may call the
Penrose process. Which, in fact one may extract energy and angular momentum
for the black hole.
Thus it is important to find the limits on the energy which a particle, at
a particular location, can have.
From the geodesic equations (\ref{radial}), we have
$$
E^2\left[(r^2+n^2)^2+a^2(2Mr-Q^2+2n^2)+a^2(r^2+n^2)\right]-2aEL(2Mr-Q^2+2n^2)
$$
\begin{eqnarray}
-L^2(r^2-2Mr-n^2+Q^2)+\epsilon \Delta (r^2+n^2) &=& 0 ~.\label{radpp}
\end{eqnarray}
Since this equation implies that there is no contribution to $E$ from the
kinetic energy derived from $\dot{r}^2$. Solving equation (\ref{radpp}) for
$E$ and $L$, separately, we obtain
\begin{eqnarray}
E=\frac{[aL(2Mr+2n^2-Q^2)\pm Y_{1} \sqrt{\Delta}]}{[(r^2+n^2)^2+a^2(r^2+2Mr-Q^2+3n^2)]} ~.\label{engpp}\\
\mbox{where} \nonumber\\
Y_{1}=\sqrt{\{L^2(r^2+n^2)^2-[(r^2+n^2)^2+a^2(r^2+2Mr-Q^2+3n^2)]\epsilon(r^2+n^2)\}} \nonumber
\end{eqnarray}
and
\begin{eqnarray}
L &=& \frac{[-aE(2Mr+2n^2-Q^2)\pm Y_{2} \sqrt{\Delta} ]}{[(r^2-2Mr+Q^2-n^2)]} ~.\label{angpp}\\
\mbox{where} \nonumber\\
Y_{2} &=& \sqrt{\{E^2(r^2+n^2)^2+(r^2-2Mr+Q^2-n^2)\epsilon(r^2+n^2)\}}
\end{eqnarray}
To derive these above equations, we have made use of the following identity:
$$
\Delta(r^2+n^2)^2-a^2(2Mr -Q^2+2n^2)^{2}=
$$
\begin{eqnarray}
[(r^2+n^2)^2+a^2(r^2+2Mr-Q^2+3n^2)](r^2-2Mr+Q^2-n^2) ~.\label{identitypp}
\end{eqnarray}
From Eq. (\ref{engpp}) we can inferred that the situations under which
$E$ can be negative, as perceived by an observer at infinity.
Firstly, it is important to observe that a particle of unit mass, at rest
at infinity, must in accordance with our conventions, be assigned an energy
$E=1$ and to be consistent with this requirement. So in the present context,
we choose the positive sign on the right hand side of the Eq. (\ref{engpp}).
It is clearly necessary that for $E<0$, $L<0$.
and
$$
a^{2}L^{2} (2Mr-Q^2+2n^2)^2>
$$
\begin{eqnarray}
\Delta (r^2+n^2)\left[L^2(r^2+n^2)-\{(r^2+n^2)+a^2(r^2+2Mr-Q^2+3n^2)\}\epsilon \right] ~.\label{ien}
\end{eqnarray}
with the aid of the identity  (\ref{identitypp}), this inequality can be brought
to the form
$$
[(r^2+n^2)^2+a^2(r^2+2Mr-Q^2+3n^2)] \times
$$
\begin{eqnarray}
[(r^2-2Mr-n^2+Q^2)L^2-\epsilon \Delta (r^2+n^2)] < 0 ~.\label{idenpp}
\end{eqnarray}
It immediately follows that $E<0$ if and only if $L<0$.
and
\begin{eqnarray}
\frac{(r^2-2Mr+Q^2-n^2)}{r^2+n^2}< \frac{\Delta}{L^2} \epsilon
\end{eqnarray}
Thus we may conclude only counter rotating particles can have negative energy;
and, on the equatorial plane, it is further necessary that $r<a+M$ i.e.,
the particle be inside the ergo-sphere. Whereas the ergo-sphere for KNTN
space-time could be found in Eq. (\ref{ergokntn}). At the extremal limit the 
ergo-sphere is situated at $r_{ergo}=M+a\sin\theta$.

\subsection{The original Penrose Process:}
In this typical process a particle, at rest at infinity, arrives by a
geodesic in the equatorial plane, at a point $r<a+M$ when it has
a turning point (so that $\dot{r}=0$). At $r$, it disintegrates into
two photons, one of which crosses the outer horizon and is lost while the
other escapes to infinity. We arrange that the photon which crosses the
event horizon has negative energy and the photon which escapes to
infinity has an energy in excess of the particle which arrived
from infinity.
Let
\begin{eqnarray}
E^{(x)}=1,\, L^{(x)}; \, E^{(y)},\, L^{(y)};\,\, \mbox{and}\,\, E^{(z)}, L^{(z)}
~.\label{e1l1e2l2}
\end{eqnarray}
denote the energies and the angular momentum of the particle arriving from
infinity and of the photons which cross the event horizon  and escape
to infinity, respectively.
Since the particles from infinity arrives at $r$ by a time-like geodesics
and has a turning point at $r$, its angular momentum, $L^{(x)}$ , can be
inferred from equation (\ref{angpp}) be setting $\epsilon=-1,E=1$,
Thus one obtains,
\begin{eqnarray}
L^{(x)} &=& \frac{[-a(2Mr+2n^2-Q^2)+\sqrt{\Delta}\sqrt{(r^2+n^2)}+
\sqrt{2Mr-Q^2+2n^2}]}{[(r^2-2Mr+Q^2-n^2)]} \nonumber\\
        &=& \alpha^{(x)} \, \mbox{(say)}
~.\label{angppx}
\end{eqnarray}
Analogously, by setting $\epsilon=0$ and choosing, respectively,
the negative and the positive sign in equation (\ref{angpp}),
we can obtain the relations between the energies and the angular
momentum of the photon which crosses the event horizon and the
photon which escapes to infinity. We get
\begin{eqnarray}
L^{(y)} &=& \frac{[-a(2Mr+2n^2-Q^2)E^{(y)}-\sqrt{\Delta}(r^2+n^2)E^{(y)}]}{[(r^2-2Mr+Q^2-n^2)]}\nonumber\\
&=& \alpha^{(y)} E^{(y)}\, \mbox{(say)} ~.\label{angppy}
\end{eqnarray}
and
\begin{eqnarray}
L^{(z)} &=& \frac{[-a(2Mr+2n^2-Q^2)E^{(z)}-\sqrt{\Delta}(r^2+n^2)E^{(z)}]}{[(r^2-2Mr+Q^2-n^2)]}\nonumber\\
&=& \alpha^{z} E^{(z)}\, \mbox{(say)} ~.\label{angppz}
\end{eqnarray}
The conservation of energy and angular momentum now requires that
\begin{eqnarray}
E^{(y)}+E^{(z)}=E^{(x)}=1
\end{eqnarray}
and
\begin{eqnarray}
L^{(y)}+L^{(z)}=\alpha^{(y)} E^{(y)}+\alpha^{(z)} E^{(z)} =L^{(x)}=\alpha^{(x)}
\end{eqnarray}
Solving these equations, we get
\begin{eqnarray}
E^{(y)} &=& \frac{\alpha^{(x)} -\alpha^{(z)}}{\alpha^{(y)}-\alpha^{(z)}}
\end{eqnarray}
and
\begin{eqnarray}
E^{(z)} &=& \frac{\alpha^{(y)} -\alpha^{(x)}}{\alpha^{(y)}-\alpha^{(z)}}
\end{eqnarray}
or, substituting for $\alpha^{(x)}$, $\alpha^{(y)}$, and $\alpha^{(z)}$
from equations (\ref{angppx}),(\ref{angppz}), we get
\begin{eqnarray}
E^{(y)} &=& -\frac{1}{2}\left(\frac{\sqrt{2Mr-Q^2+2n^2}}{r^2+n^2}-1 \right)
\end{eqnarray}
and
\begin{eqnarray}
E^{(z)} &=& +\frac{1}{2}\left(\frac{\sqrt{2Mr-Q^2+2n^2}}{r^2+n^2}-1 \right)
\end{eqnarray}
The photon escaping to infinity has, indeed, an energy in excess of
$E^{(x)}=1$ so long as $r<a+M$ (as we have postulated). The energy,
$\Delta E$, that has been gained is
\begin{eqnarray}
\Delta E &=& \frac{1}{2}\left(\frac{\sqrt{2Mr-Q^2+2n^2}}{r^2+n^2}-1 \right)=-E^{(x)}
~.\label{chaeng}
\end{eqnarray}
It is observed that from Eq. (\ref{chaeng}) by the Penrose process the maximum
gain in energy that can be achieved is when the particle, arriving from infinity, has a
turning point at the event horizon . Thus
\begin{eqnarray}
\Delta E \leq \frac{1}{2}\left(\sqrt{1+\frac{a^2}{r_{+}^{2}+n^2}}-1\right)
~.\label{chaeng1}
\end{eqnarray}

\begin{figure}
\begin{center}
{\includegraphics[width=0.45\textwidth]{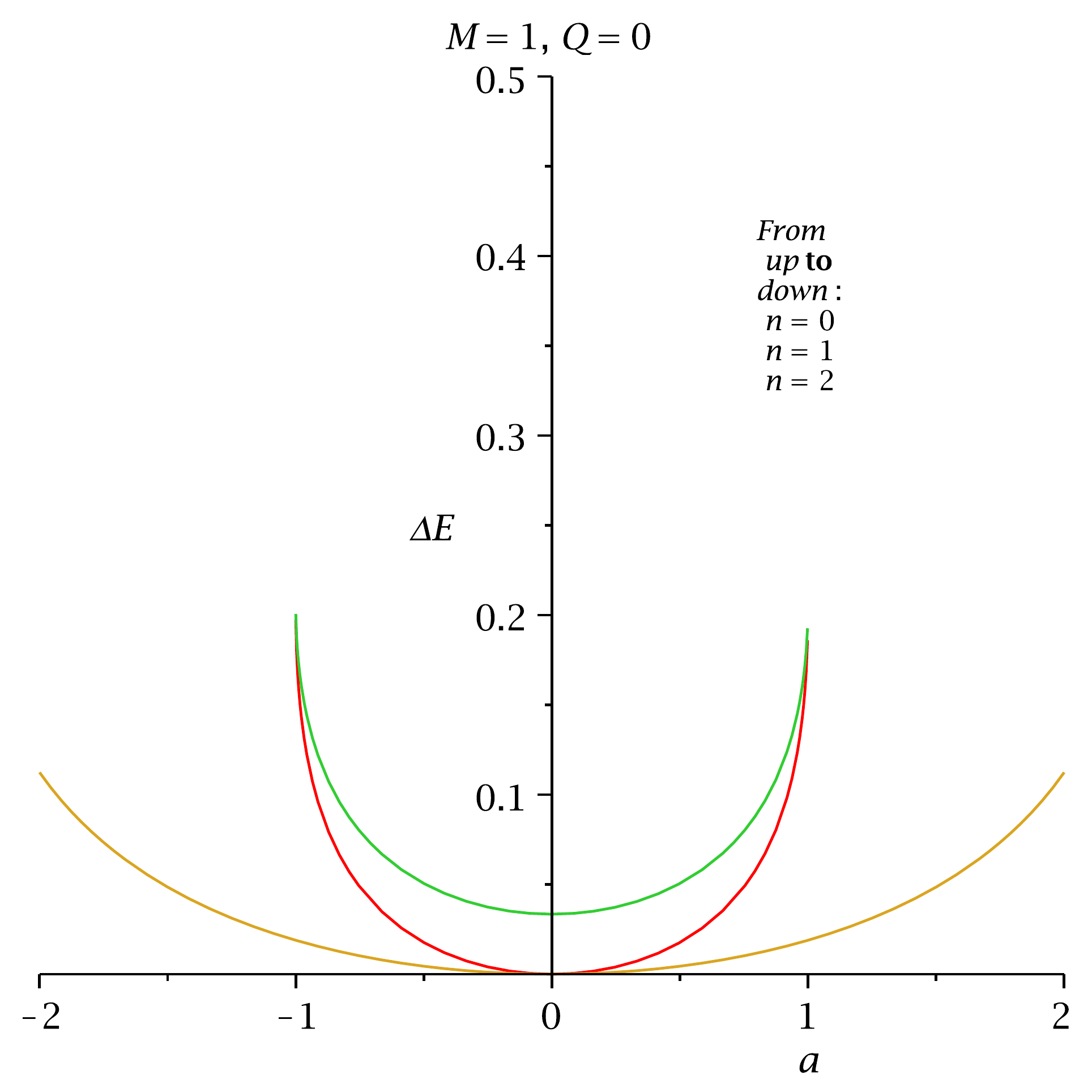}}
{\includegraphics[width=0.45\textwidth]{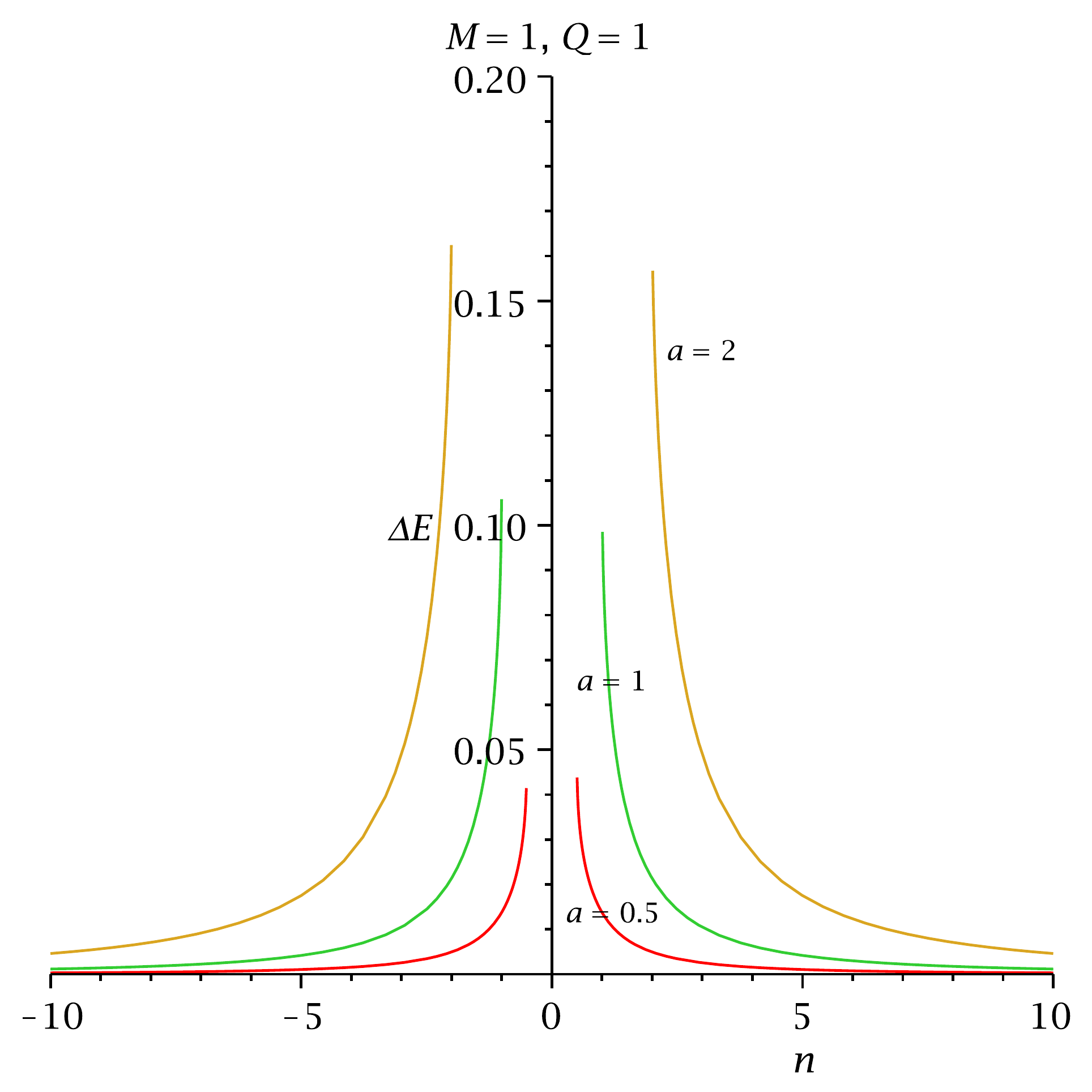}}
\end{center}
\caption{The figure shows the variation  of $\Delta E$  with $a$ and $n$ for KTN
and KNTN black hole.
\label{vf}}
\end{figure}
%

%
%

%
%
%
%

%
%

In the limit $n=Q=0$, we obtain the maximum gain in energy by extremal
Kerr black hole  which is
\begin{eqnarray}
\Delta E \leq \frac{\sqrt{2}-1}{2}=0.207
\end{eqnarray}
This can be seen from the left side of Fig.\ref{vf}. The right side of Fig.\ref{vf}
implies that when the value of NUT parameter increases the energy gain decreases.
The 3D view of the energy variation could be seen from the Fig.\ref{3de}, Fig.\ref{3de1},
Fig.\ref{3de2}, Fig.\ref{3de3}.

It is also found that from Eq.\ref{chaeng1}, the gain in energy by
Penrose Process depends on the both charge $(Q)$ and NUT parameter $(n)$.
It is important to note that one can obtain easily all the above equation
for KTN black hole when taking the limit $Q=0$.

\section{\label{dis} Discussion:}

In this work,  we have described a detailed analysis of the geodesic motion
of both massive particles  and massless particles in the space-time  of
KNTN black hole which is the most general stationary, axially symmetric,
non-asymptotically flat electro-vac space-time. It is therefore characterized by
four parameters, namely mass, spin, electric charge and the NUT parameter.
The special characteristics of this black hole is that
it is of Petrov-Pirani type-D and the photon trajectories are doubly degenerate
principal null congruences. The other feature is that the geodesic equations
are separable in Boyer-Lindquist type coordinates.

We have derived the conditions for the existence of ISCO, MBCO and CPO of the
said black holes.  We also computed some other important astronomical quantities
like ($L_{0}, E_{0}, v^{\phi},\Omega_{0}, T_{0}$) for KNTN black hole which are relevant
for accretion disk theory. The study of effective potential implies that
due to the presence of the NUT parameter the shape of the potential barrier
get modified in contrast with zero NUT parameter. We have further studied the
Penrose process of KNTN black hole. It is found that the gain in energy by the
Penrose Process explicitly depends on the both charge and NUT parameter
of the space-time.  It is shown that the presence of the NUT parameter may affects
in the energy extraction process. When NUT parameter is increasing the gain in
energy is decreasing.

Since we have restricted to the black hole space-time here, so as an extension of
this work, it would be interesting to investigate the detailed analysis of the
properties of circular geodesics for the naked singularity cases in comparison
with the black hole cases for the above mentioned  space-time. For
KN black hole \cite{ruffini}, the authors showed the presence
of a typical band structure  i.e. a disconnected region of
stable orbits for the space-time generated by KN naked singularity.
This band structure is completely absent in the case of black holes.

\section*{Acknowledgements}
The author is grateful to the Inter-University Centre for Astronomy and Astrophysics(IUCAA), Pune
for hospitality during the final stages of this work.

\end{document}